\documentclass[a4paper,fleqn]{cas-sc}
\usepackage{setspace}
\usepackage[switch]{lineno}
\usepackage[numbers]{natbib}
\usepackage{nccmath}
\usepackage{lineno}
\usepackage{float}
\usepackage{amsmath}
\usepackage{parskip}
\usepackage{caption} 
\usepackage{upgreek} 
\usepackage[section]{placeins}
\usepackage{subcaption}
\usepackage{makecell}
\usepackage{textcomp}
\usepackage[utf8x]{inputenc}
\usepackage{siunitx}

\begin{document}

\let\WriteBookmarks\relax
\def\floatpagepagefraction{1}
\def\textpagefraction{.001}
\shorttitle{}
\shortauthors{Dylan, Mamta, Aseet, Klara et~al.}

\title [mode = title]{Photogrammetry for Precise, Rapid Module Alignment in the Mu2e Tracker}                
\author[1]{D. Palo}[orcid=0000-0001-9256-5348]
\cormark[1]
\ead{dylan.c.palo@gmail.com}
\author[1]{M. Jangra}[orcid=0000-0002-7766-1718]
\author[1]{A. Mukherjee}[orcid=0000-0001-5151-4333]
\author[1]{K. Northrup}

\address[1]{Fermilab}


%\linenumbers

\begin{abstract}
We discuss a photogrammetry technique to measure the inter-module alignment in the Mu2e tracker. The tracker consists of 216 modules each with 96 straw tubes. The relative alignment of the straw tubes in a module was previously measured by an X-ray technique, which tied the straw tube measurements to three fiducials per module. The photogrammetry technique uses an array of 15 cameras to make precise position measurements of the module fiducials in a global tracker coordinate system. Measurements of the fiducial position and radius on the camera's CCD yield $\sim\hspace{-0.1cm} \SI{25}{\micro\meter}$ resolution in the transverse directions and $< \SI{300}{\micro\meter} $ along the camera axis. In addition, we describe a procedure to combine the images with "local" mechanical measurements to yield $<100 \unit{\micro\meter}$ precision along the camera axis. The technique offers a touch-less, rapid (\textasciitilde{} 1 hour), and affordable metrology approach within the requirements of the tracker.

\end{abstract}

\maketitle

\singlespacing
%\linenumbers
\section{Introduction}

Many modern particle physics experiments rely on precise measurements of particle kinematic variables to advance the experimental state of the art. To achieve this high precision, one must determine the relative alignment of the sub-components in a detector to high precision. Without this alignment, one creates systematics in the reconstructed kinematic variables thus degrading measurement precision and ultimately experimental sensitivity.

In the Mu2e experiment, a low-mass straw tube tracker must precisely determine the momentum of electrons associated with the coherent conversion of a muon into an electron. A precise momentum measurement is required to distinguish the signal from Standard Model background\cite{Mu2e}. Precise alignment of the tracker straw tubes is required for this momentum measurement. 

The tracker consist of 216 modules each with 96 straws tubes. The relative alignment of the straws tubes in a module was previously measured with respect to a set of 3 fiducials using an X-ray technique\cite{OH201664}.  By imaging the same fiducials measured by the X-ray scan, we transform the single module measurements into a global tracker coordinate system. In this paper, we discuss a photogrammetry technique used to determine the module-to-module alignment in the straw tube tracker.

The photogrammetry technique offers an affordable (O($\$10k$), rapid O(1 hour) determination of the positioning of the modules in the Mu2e tracker. The motivation behind the photogrammetry approach is described in Section \ref{sect:Motivation}. The details of this photogrammetry technique were based on the Mu2e tracker design, but the technique could be used more generally when building a detector with many sub-components.

\subsection{Mu2e Experiment}
\label{sect:Mu2e}
The Mu2e Experiment at Fermilab is an upcoming search for the $\mu^{-} N \rightarrow e^{-} N$ conversion process. The conversion is an instance of charged lepton flavor violation where the signal is a monoenergetic electron with momentum of the muon rest mass minus the nucleus recoil energy and the binding energy (104.97 MeV/c).

A total of $\sim 6.7\cdot 10^{17}$\cite{Mu2e:2014fns} muons will stop on aluminum foils. The vast majority of these muons will either be captured in the Aluminum target nuclei (61\%) or decay in the orbit (DIO) of Aluminum nuclei resulting in an electron and two neutrinos. One of the key backgrounds comes from instances where the muon exchanges momentum with the nucleus during the muon DIO. In this case, the DIO electron momentum can approach the muon rest mass. The experiment relies on a precision measurement of the electron momentum to distinguish between a high-energy DIO electron and the signal.

This electron momentum measurement is performed by the Mu2e tracker. The tracker measures the electron position at many points along its helical trajectory in a well-known magnetic field, thus measuring the electron momentum. 

\subsection{Mu2e Tracker}

\begin{figure*}[t]
    \centering
    \begin{subfigure}[t]{0.25\textwidth}
        \centering
        \includegraphics[height=1.7in]{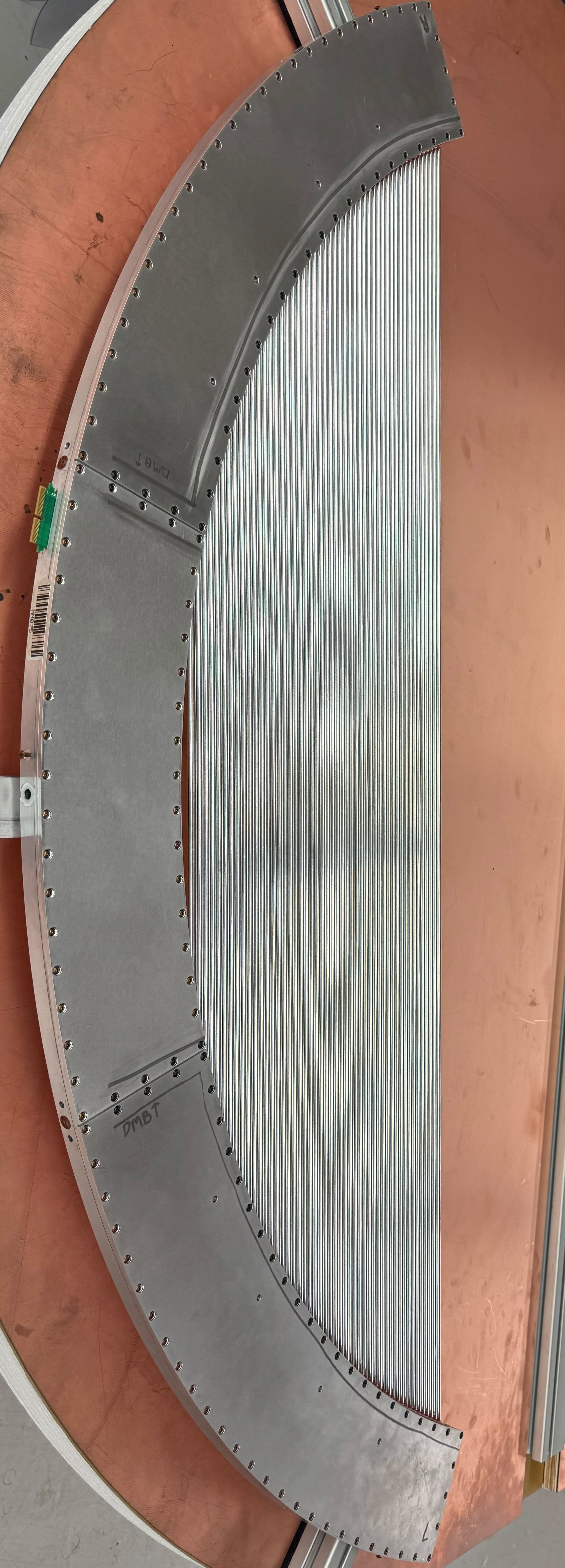}
        \caption{}
    \end{subfigure}
    \begin{subfigure}[t]{0.26\textwidth}
        \centering
        \includegraphics[width=\textwidth]{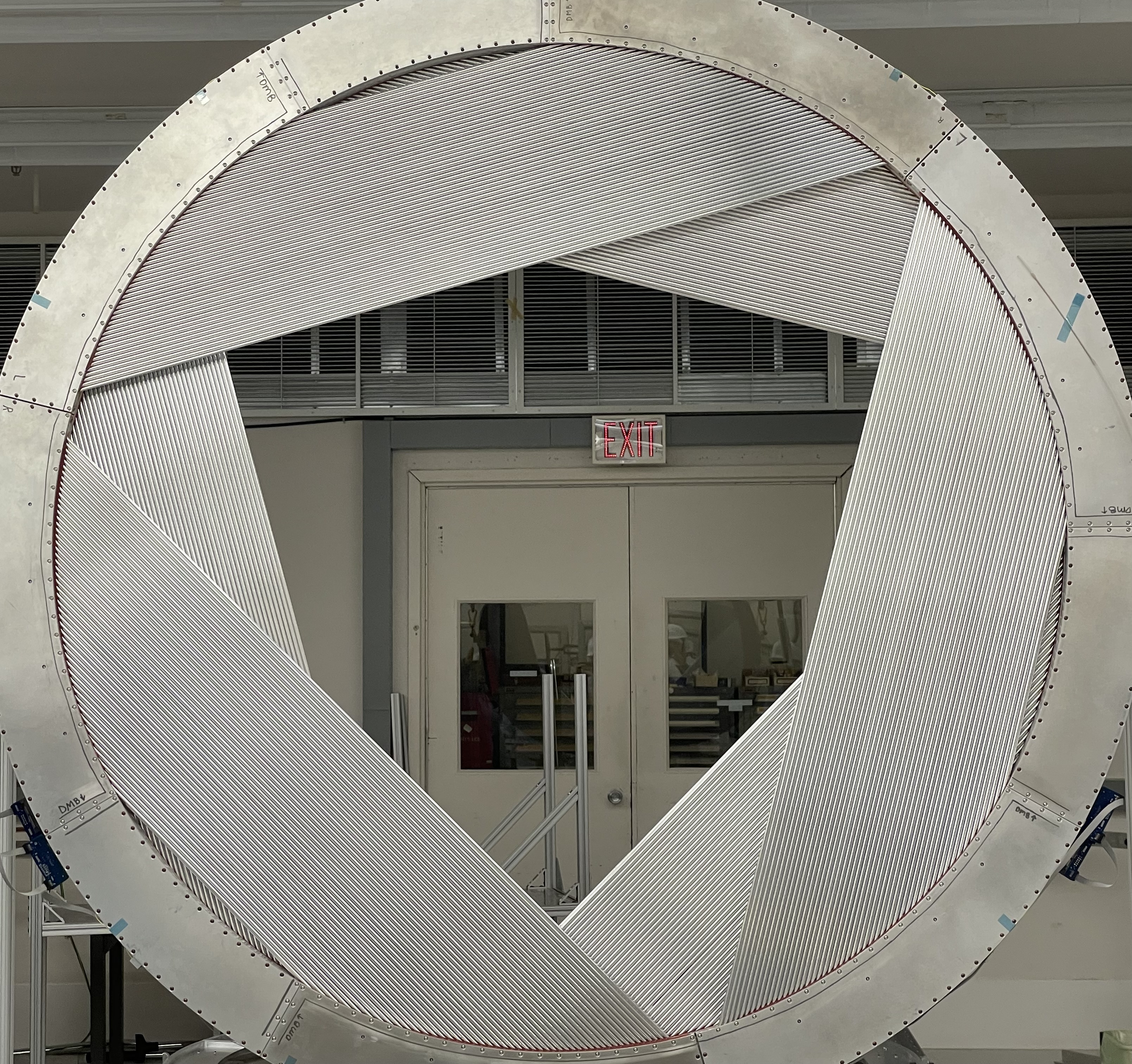}
        \caption{}
    \end{subfigure}
    \begin{subfigure}[t]{0.25\textwidth}
        \centering
        \includegraphics[width=\textwidth]{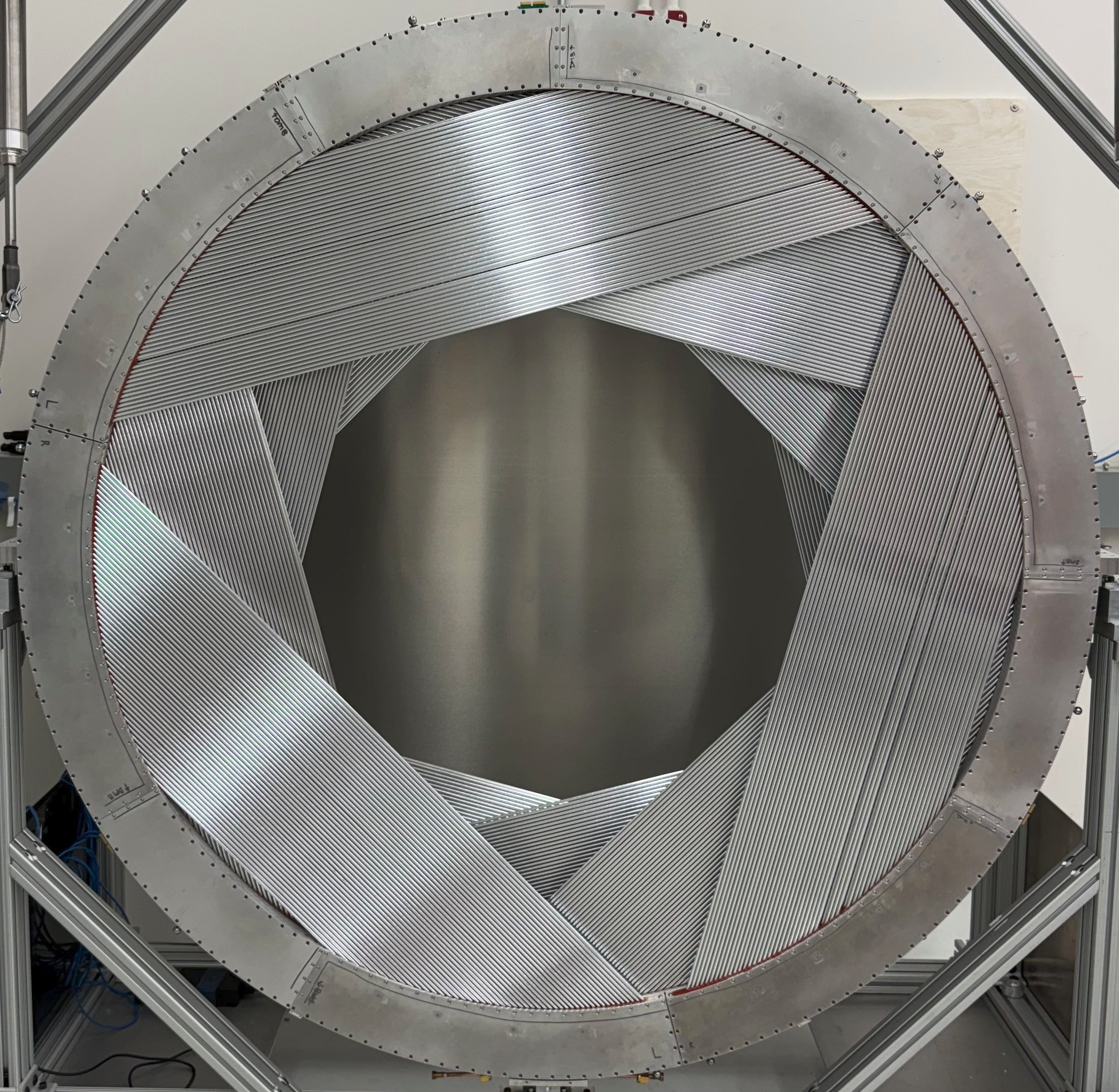}
        \caption{}
    \end{subfigure}
    \begin{subfigure}[t]{0.25\textwidth}
        \centering
        \includegraphics[width=\textwidth]{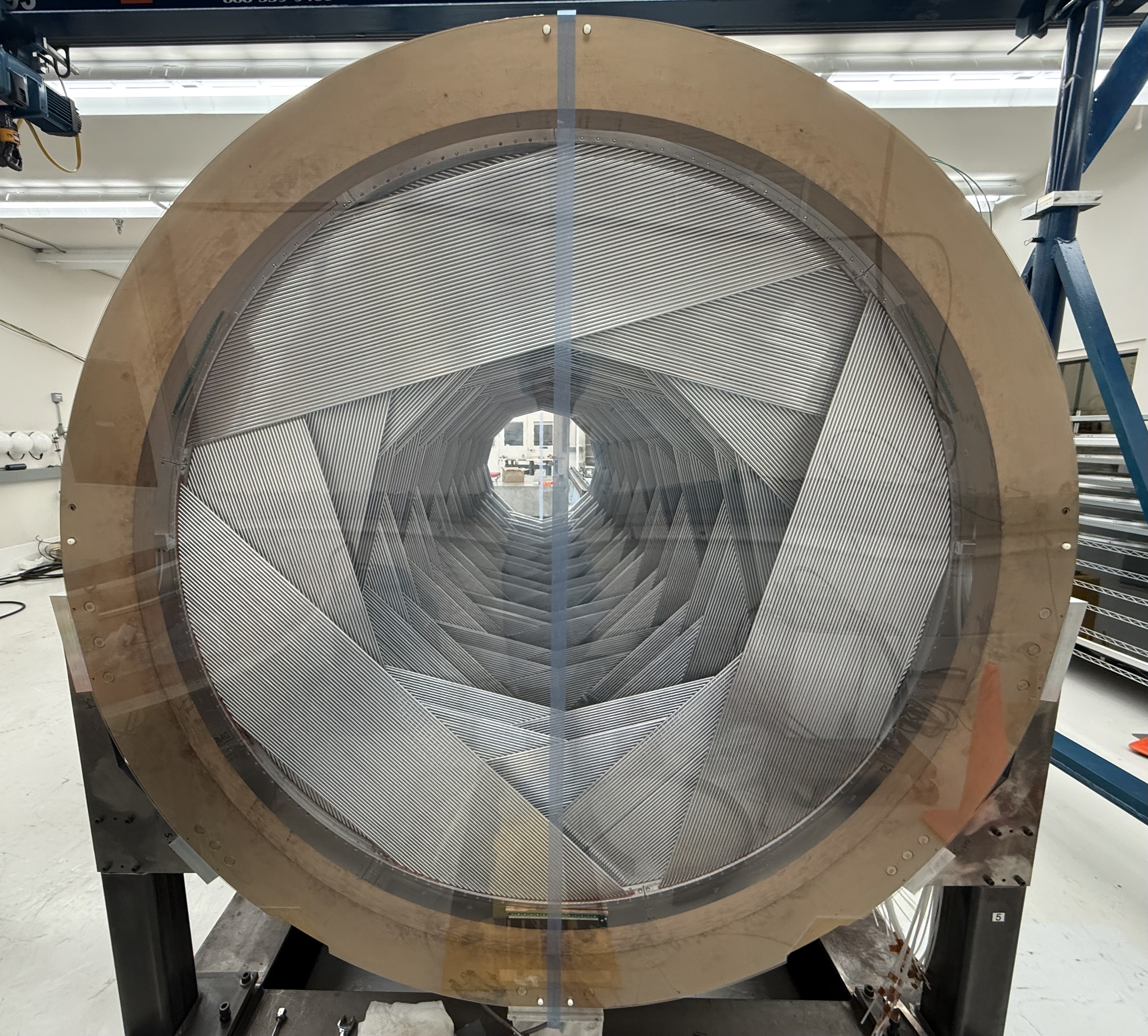}
        \caption{}
    \end{subfigure}
        \begin{subfigure}[t]{0.5\textwidth}
        \centering
        \includegraphics[width=\textwidth]{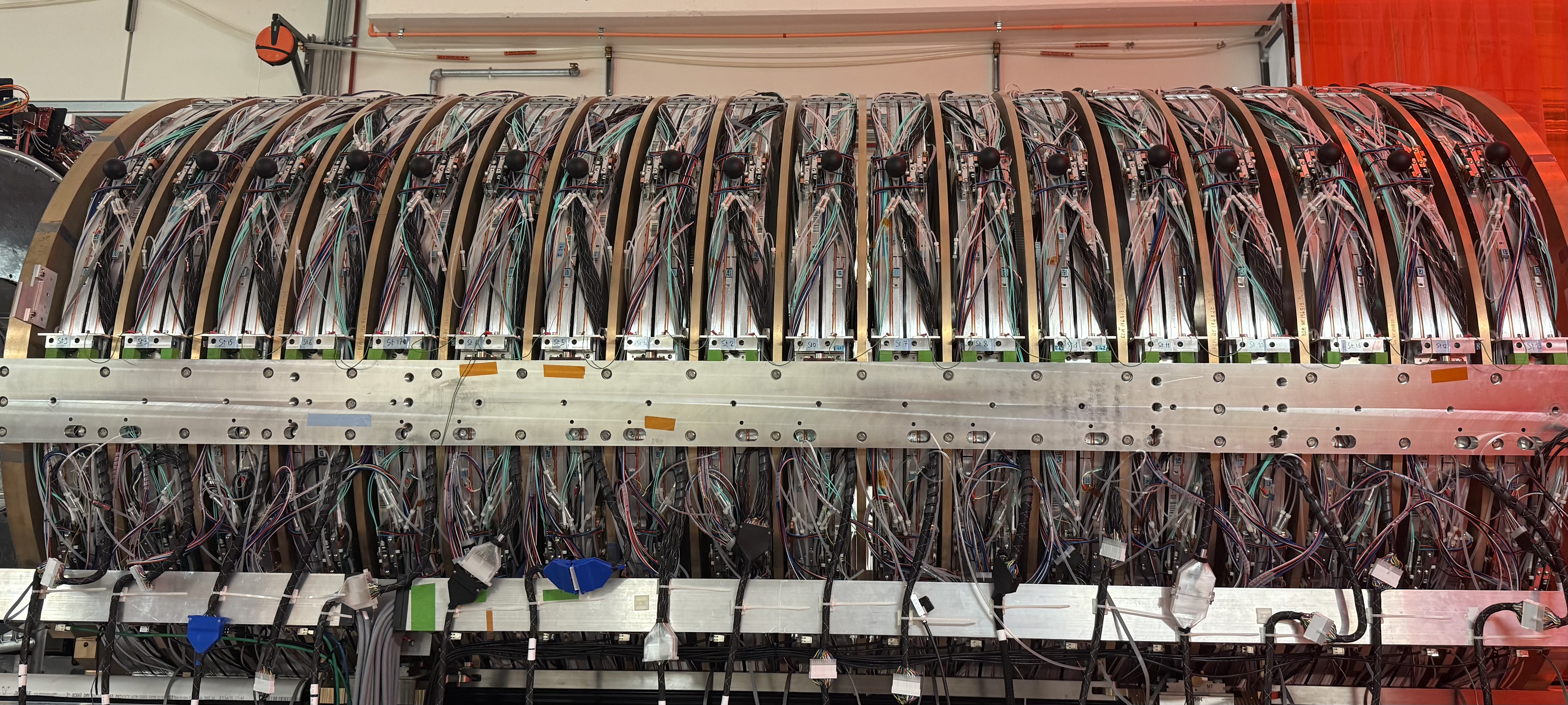}
        \caption{}
            \label{TrackerFrame}

    \end{subfigure}
    \caption{Figures of the sub-modules in the Mu2e tracker. (a) Tracker panel (96 straw tubes). This is the smallest module in the tracker. (b) Tracker plane (6 panels). (c) Tracker station (2 planes). (d) Fully assembled tracker in frame. In this view, +X is to the left, +Y is up and +Z is into the page.  (e) View along X showing the $Z_{Mu2e}$ span of the fully assembled tracker in frame. Stations are spaced between brass rings. }

    \label{TrackerLayout}
\end{figure*}

\label{sect:tracker}
The tracker consists of 20,736 5-$\unit{\milli\meter}$ diameter, 15-$\unit{\micro\meter}$-thick aluminized Mylar straw tubes. The straw tubes include a 25 $\unit{\micro\meter}$ gold-plated tungsten sense wire.

The tracker has a modular design (Figure \ref{TrackerLayout}). First, 96 straw tubes are assembled into 'panels', the smallest module in the tracker. The panels are assembled into circular 'planes' (36 planes total) with two layers of 3 panels each. Pairs of planes are then joined into 'stations' (18 stations total) where the two planes are rotated $180^{\circ}$ about the Y axis (vertical in Figure \ref{TrackerLayout} (b)) with respect to one another. These stations are housed in the tracker frame spaced along the beam axis. The stations have a radius of $\sim 800 \unit{\milli \meter}$ and a length along the beam axis of $\sim 3000 \unit{\milli \meter}$.

At each stage of the tracker assembly (panel assembly, plane assembly, station assembly, station-to-station alignment in the frame), variations and systematics in the alignment of the order $100s$ of $\unit{\micro\meter}$ are observed. 

The stations are positioned in the frame using three-point kinematic mounts. These kinematic mounts fully defines the rigid body position of the stations (three angles and three positions) without putting stress on the detector if it deforms (e.g. expands/contracts due to temperature). Figure \ref{TrackerFrame} shows the stations in the frame.

 The first station mount is a ball/cone connection at +X defining an XYZ position. The second is a "canoe sphere" in a V-groove at -X defining a YZ position while allowing the X position to float. The canoe sphere has a span of 3/4" ($19.05 \unit{\milli \meter}$) while containing two spherically curved sections each with a radius of 10" ($254 \unit{\milli \meter}$) to allow for two points of contact inside on the kinematic V (see \cite{BallTec} for more details). The third mount defines a Z position at -Y. The Z position is defined by a pin pressed against a slot via a spring.

Stations are assembled with the same nominal geometry except the kinematic ball and canoe sphere positioning. Alternating stations in the tracker frame have the two mounts on opposite station arms (located at +X/-X). Thus every other station in the frame is rotated by $180^{\circ}$ about the Y-axis. 

\begin{figure*}[htb]
    \centering
    \begin{subfigure}[t]{0.27\textwidth}
        \centering
        \includegraphics[width=0.9\textwidth]{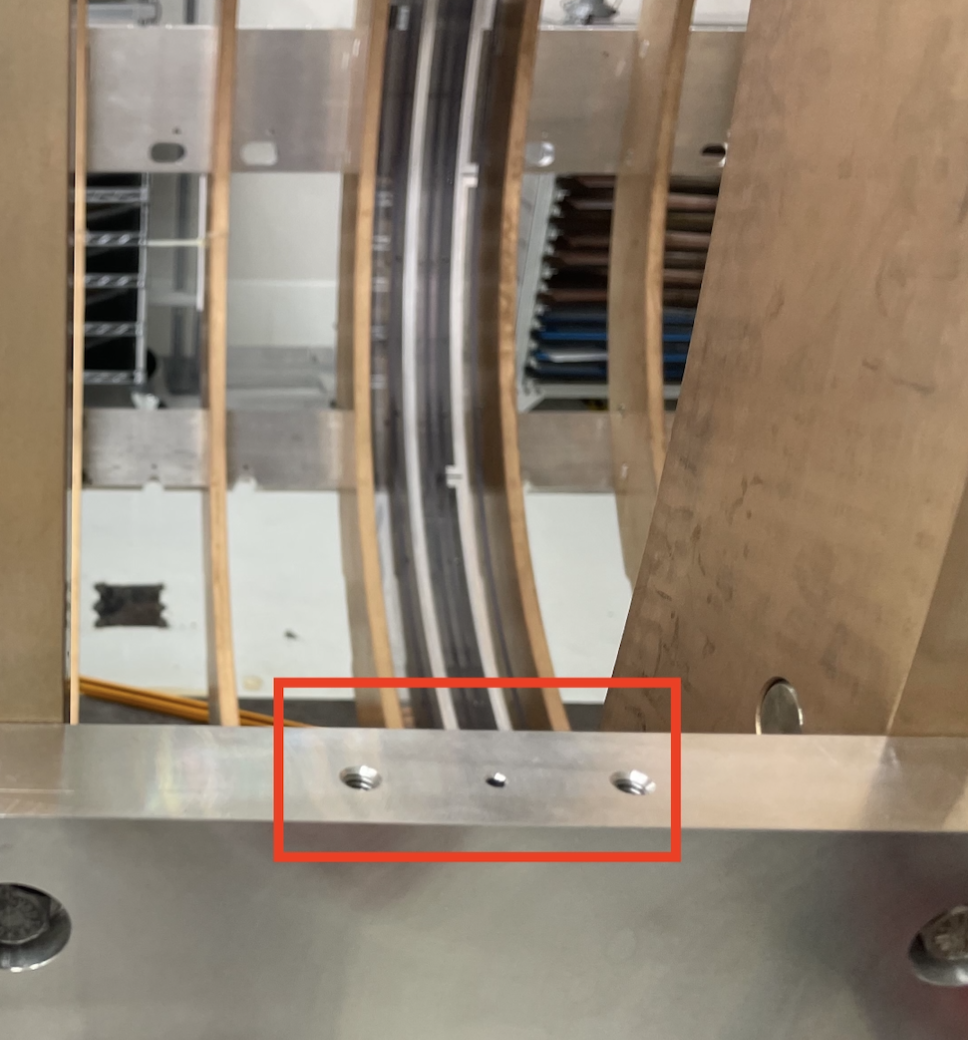}
        \caption{}
    \end{subfigure}
    \begin{subfigure}[t]{0.3\textwidth}
        \centering
        \includegraphics[width=0.9\textwidth]{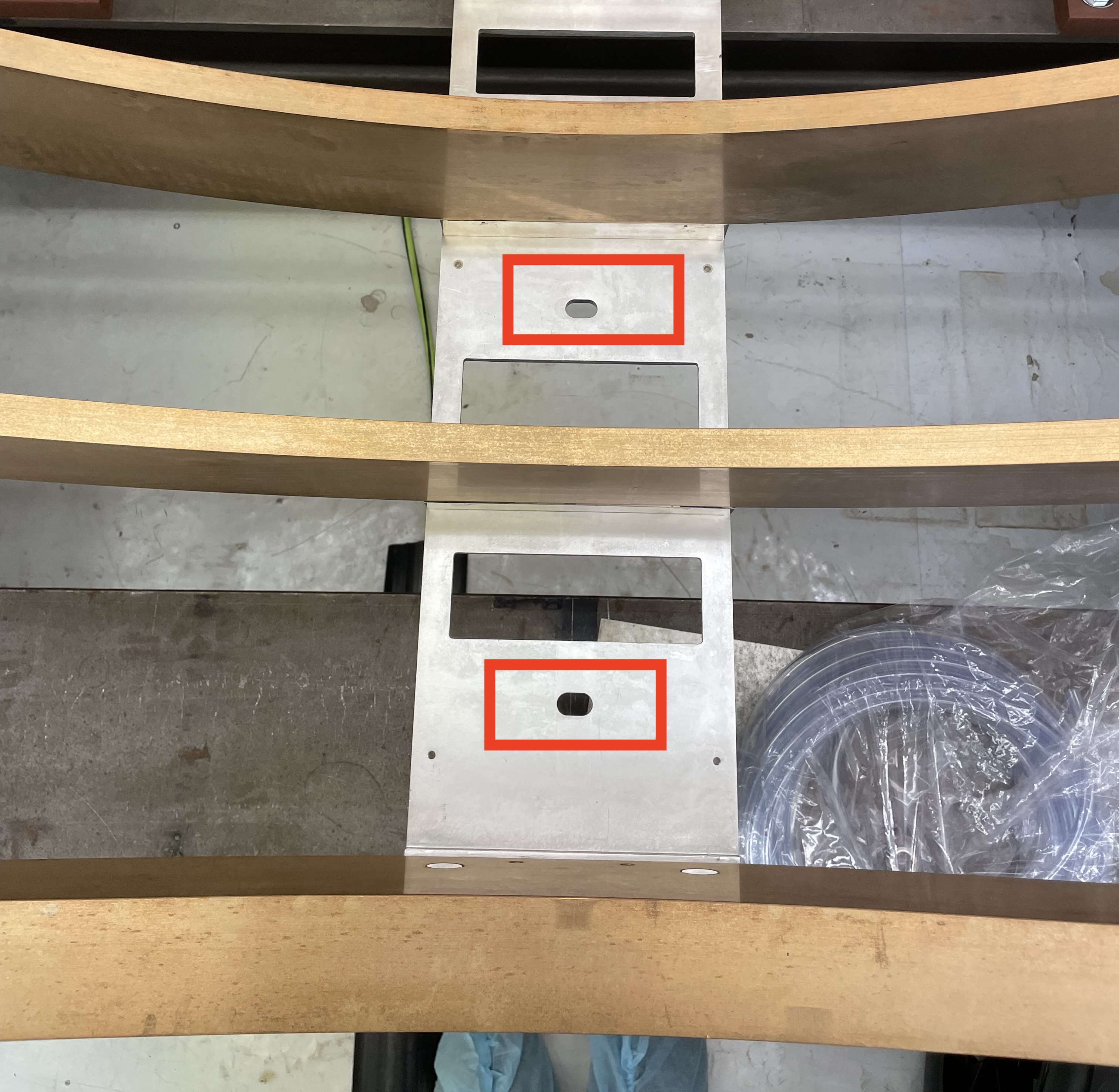}
        \caption{}
    \end{subfigure}        
    \\
    \begin{subfigure}[t]{0.3\textwidth}
        \centering
        \includegraphics[width=0.9\textwidth]{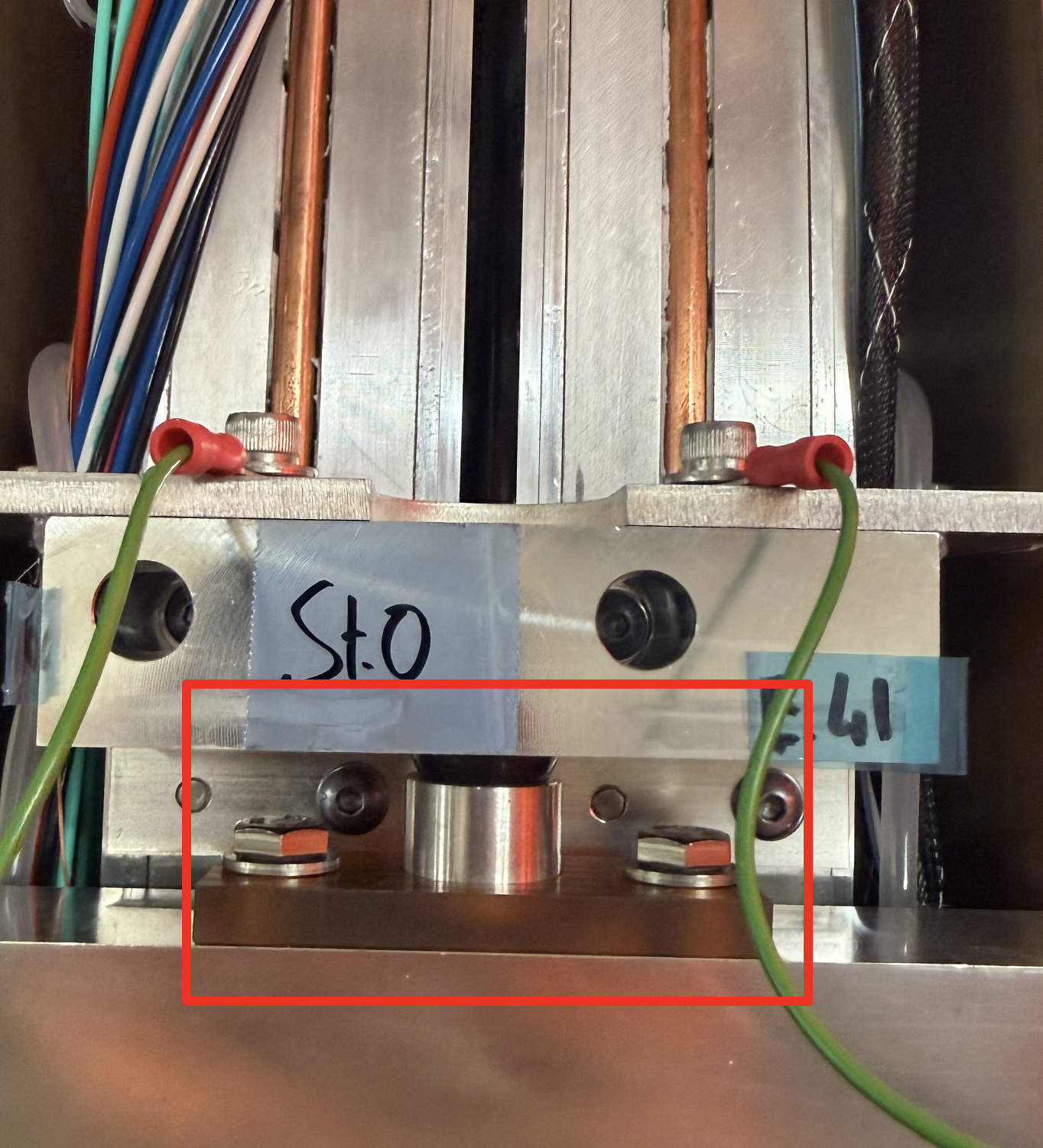}
        \caption{}
    \end{subfigure}
    \begin{subfigure}[t]{0.3\textwidth}
        \centering
        \includegraphics[width=0.9\textwidth]{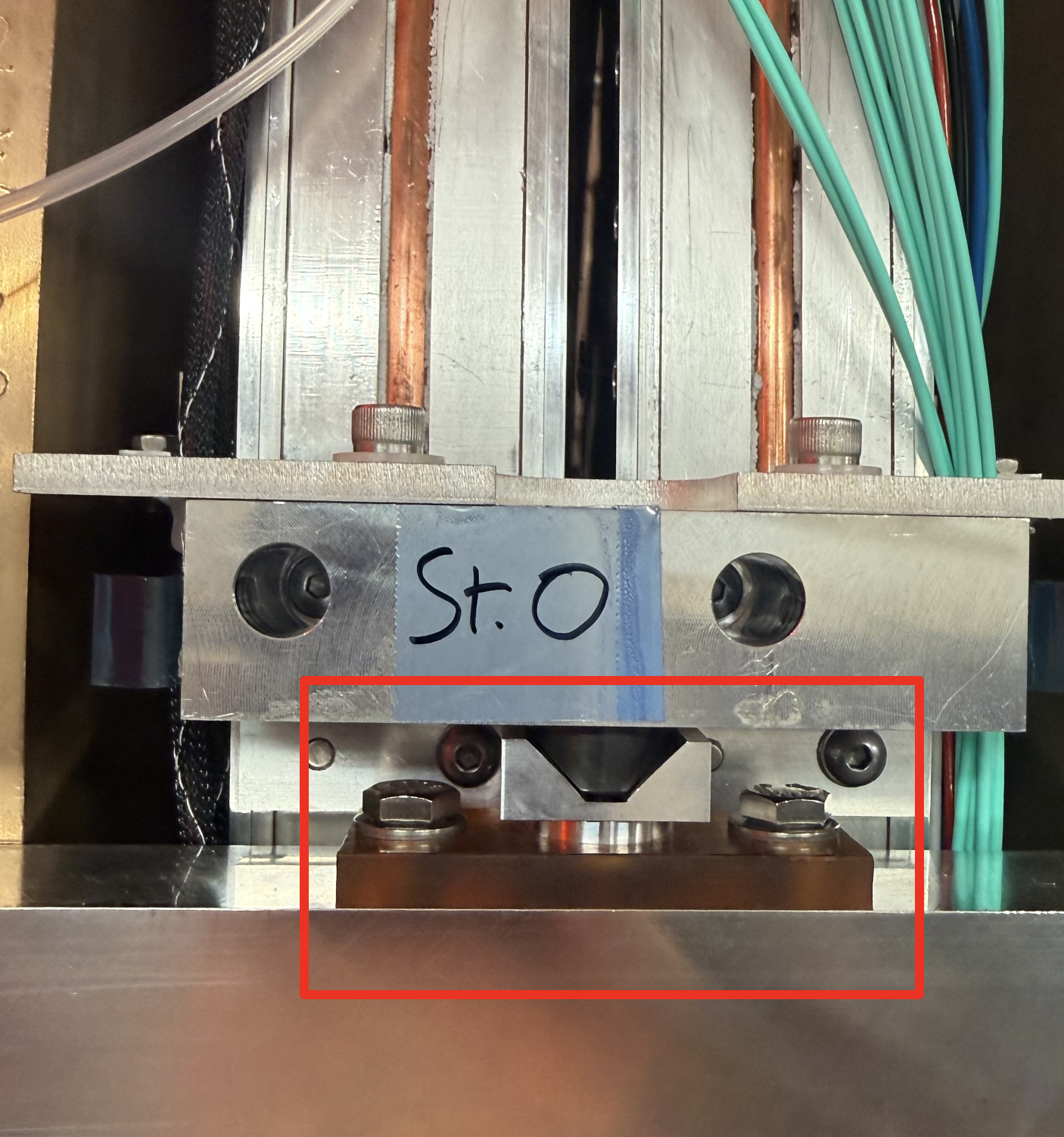}
        \caption{}
    \end{subfigure}

    \caption{Figures detail features on the tracker frame used to align the stations. (a) The 1/8" ($3.175 \unit{\milli \meter}$) pin hole on the tracker frame stave plus the pairs of threaded holes. (b) The tracker frame Z slots at -Y (smaller slots).  (c) The cone station mount engaged on the stave 1/8" ($3.175 \unit{\milli \meter}$) pin. (d) The V-groove station mount engaged on the stave 1/8" ($3.175 \unit{\milli \meter}$) pin.   }
    \label{framefids}
\end{figure*}

The position of the kinematic mounts on the frame are defined by 1/8" ($3.175 \unit{\milli \meter}$) pin holes (located at +X and -X) on the two staves and the slot on the bottom stave (located at -Y). Figure \ref{framefids} (a) shows two sets of pin holes on the stave, and Figure \ref{framefids} (b) shows the frame slots defining the Z position of the station at -Y.

 The position of the kinematic cone and V groove are machined with respect to these 1/8" ($3.175 \unit{\milli \meter}$) pin holes; an 1/8" pin engages both the kinematic mount and the pin hole in the frame. The kinematic mounts at +X and -X are each screwed in using the 2 threaded holes per pin on the staves. Figure \ref{framefids} (c) and (d) show the cone and the V-groove engaged in the pin hole on the stave.

\subsection{Tracker Alignment Motivation}
Precise and accurate wire-to-wire alignment at the $\sim 100 \unit{\micro\meter}$ level in the Mu2e tracker is motivated by two main considerations. 

First, wire-to-wire misalignments degrade the overall track resolution. High-energy DIO events reconstructed with a large positive-momentum-error can end up in the signal region. These backgrounds are reduced through quality cuts in the track selection procedure (details found in Ref. \cite{Edmonds:2021lzd}). This selection procedure requires a clear understanding of fit consistency. Therefore, the alignment goal is to not significantly degrade the intrinsic detector resolution and therefore the track fit consistency. With an alignment goal of $100 \unit{\micro \meter}$, the contribution of residual misalignment to resolution will be considerably less than the goal intrinsic hit resolution of $160 \unit{\micro \meter}$.

In addition, errors in the momentum scale change the number of DIO events in the signal region. Shifting the absolute momentum scale by $\pm100$ keV/c shifts the DIO background estimate asymmetrically by $[+59\%, -37\%]$; details are available in Ref. \cite{Mu2e}. During standard Mu2e data-taking there is no direct kinematic line or boundary to calibrate the momentum scale (e.g. $E_{e^{+}} < 52.83 MeV/c^{2}$ for Michel decay positrons\cite{MEGII}). 

The momentum scale uncertainty is the accumulation of systematics from the magnetic field mapping, the straw tube sense wire positioning, and any unaccounted for material effects. The overall goal is to determine the momentum scale to better than $\pm100 keV/c$ i.e. a part in $\sim10^{3}$.  With a maximum active tracker radius of $680 \unit{\milli \meter}$, achieving an alignment precision and accuracy of $\sim 0.1 mm$ would suppress contributions to the momentum scale error to $O(10^{-4})$. 

These two DIO background considerations motivate a wire-to-wire alignment to a precision and accuracy of $100 \unit{\micro \meter}$.

\subsection{Photogrammetry Motivation}
\label{sect:Motivation}

Traditionally, one would use a Coordinate Measuring Machine (CMM), FARO arm\cite{FAROQuantumX}, or a laser tracker (e.g.\cite{LaserTracker}) to make precise position measurements. There were several reasons these options were suboptimal. A few requirements for the metrology are bulleted below: 

\begin{itemize}
    \item The measurement should be done \textit{before} the stations are mounted in the frame to allow us to correct station-to-station variations into custom station kinematic mounts. 
    \item The tracker work requires a clean room; the metrology would be ideally performed in a clean room. 
    \item The procedure should avoid tools being above the center of the tracker to avoid tools falling into the straws. Further, the procedure should minimize handling of the stations to avoid damage to the straws. 
    \item The metrology should be quick to avoid delays in the station assembly, cabling, insertion of stations into the frame, etc.
    \item The metrology procedure should be affordable; this includes the person-hours to operate the equipment. 
    \item The metrology precision should be better than $100 \unit{\micro\meter}$ in all directions. 
\end{itemize}

By making the measurements \textit{before} the stations are mounted in the frame, we can measure station-to-station variations. We can later offset each set of station kinematic mounts to account for these station-to-station variations in the frame. This is described in Section \ref{StationOptimization}. 

Fermilab has a large CMM that was used to measure the tracker frame. This would have difficulties accessing all of the panel fiducials (e.g. those at the bottom) without modifications to the CMM probes. In addition, this device cannot be moved so this would require additional shipping and handling of the tracker. 

Second, one could envision using a FARO arm. This would not be able to measure all the fiducials in the frame due to the FARO arm reach, but one could imagine surveying each station separately. However, tests with the FARO suggested the time per station would be significantly more than a day; the device itself is also expensive and requires expertise. 

Third, one could envision using a laser tracker. It would still be difficult to measure the stations in the frame using a laser tracker due to the obstructing staves, cabling, etc., but one could imagine surveying each station in a dedicated stand before the stations are inserted in the frame. One concern was that the laser tracker requires moving a probe to each fiducial, thus requiring handling above the stations. In addition, tests using our \textit{calibration station} (Section \ref{sect:calibrationstation}) suggested this would require about a day per station. The significant amount of time per station is due to the limited space in the tracker clean room. The laser tracker measurement requires moving the device to several locations to survey all fiducials. Further, this is an expensive device and requires expertise. 

 The photogrammetry technique offered a touch-less, rapid, and affordable approach. In addition, our initial tests with a single camera on a translation stage suggested the camera would achieve precise measurements perpendicular to the camera axis ($\sim 25 \unit{\micro\meter}$) and $< 100 \unit{\micro\meter}$ resolution along the camera axis (this will be described in more detail in Section \ref{sect:camAxis}).

\section{Methods}

\subsection{Overview}
In this section, we detail the photogrammetry technique and other relevant tracker metrology. 

The photogrammetry technique optically measures the position of tracker panel fiducials after the panels are assembled into stations. These optically measured positions should mimic those measured in individual single-panel X-ray scan measurements. This X-ray technique measured the position of the straws and wires in a single panel with respect to these reference fiducials. 

The stations are imaged in a custom camera stand with 15 cameras. Using image analysis and projection equations, the images are converted into 3D positions in individual camera frames. These camera frame measurements are tied to a station-wide coordinate system using a one-time laser tracker of a reference artifact. In addition, by measuring the position of reference positions on the tracker frame itself, these single-station measurements are transformed into a tracker-wide coordinate system.

The remainder of the overview provides additional information and references to the subsections. The individual panel X-ray measurements are detailed in Section \ref{sect:xray}.  Information on the panel fiducials is detailed in Section \ref{sect:fids}. 

 All the photogrammetry measurements are made in a custom camera stand; the stand is detailed in Section \ref{sect:stand} and the relevant coordinate systems are detailed in Section \ref{sect:coordinateSystem}. The stand contains 15 cameras to view the fiducials on the single mounted station. The stand includes an array of LEDs for back-lighting. Details on the operation of this photogrammetry are in Section \ref{sect:op}.

The camera metrology requires a metrological artifact, which is used as a reference for the cameras. This artifact establishes a reference position of fiducials in a station-wide frame for the camera array, and thus the transformation from individual camera frames to the station coordinate system. The artifact is called the $\it{calibration}$ station as it contains the fiducials of a production station, but with no detector components. The calibration station was mounted in the metrology stand and surveyed by a laser tracker. The calibration station and the survey are described in Section \ref{sect:calibrationstation}. 

We detail the image analysis used to determine the best fit 2D center and radius of the tooling balls in the station on the camera CCDs in Section \ref{sect:ImageAnalysis}. This is the core data used in the metrology. This involves first finding and fitting the pixels on the edge of the tooling ball contour to a circle. The center and radius are used to reconstruct the 3D position of the tooling ball in the camera frame. 

However, we found that the radius was only able to achieve a reproducible precision of $<300 \unit{\micro\meter}$ along the camera axis. Therefore we supplemented the technique with mechanical measurements of the lengths between the three imaged fiducials. By measuring the lengths precisely (shown in Figure \ref{Triplet}), the camera analysis uses these measured lengths instead of the measured radii to get a significantly more precise measurement along the camera axis ($\sim 100 \unit{\micro\meter}$). These measurements are described in Section \ref{sect:camAxis}. In Section \ref{sect:opticalProject}, we discuss the optical projections to convert the camera data and the mechanical measurements into 3D measurements in the camera frame. 

The 3D fiducial positions in individual camera frames are transformed into a station-wide frame using the survey of the calibration station. This transformation is described in Section \ref{sect:TransformationCameraLaserTracker}. We use the same transformation for each production station. 

With the fiducials measured in a station-wide frame, we describe the procedure to transform the individual panel X-ray data into the station-wide frame in Section  \ref{sect:TransformationPanelToLaserTracker}.

Finally, the fiducials and the X-ray scanned straws and wires are transformed from the metrology stand to the global tracker frame coordinate system, this process is described in Section \ref{sect:TransformationStationWideToFrame}.

\subsection{X-ray Scan}
\label{sect:xray}
Prior to the plane and station assembly, each panel was individually scanned by an X-ray source/receiver to determine the position, orientation, and sagitta of the straws and wires with respect to three fiducials on each panel\cite{OH201664}. The X-ray measurement was done twice each with a different source angle; this allows for a stereo reconstruction of the wire and straw position. 

Each panel contains three precision 0.25" ($6.35 \unit{\milli \meter}$)  holes on the radially exterior surface each with a precision spot face: HV, gas flow in, and gas flow out ports. These holes were measured using a 0.0001" ($0.00254 \unit{\milli \meter}$) pin set; of all the pin holes in the tracker (648), only 4 are larger than 0.2503" ($\sim 8 \unit{\micro \meter}$ over nominal). 

In the panel X-ray scan, custom fiducials were inserted into the panel such that the X-ray measurement resulted in a 3D fiducial position intended to match the position of a standard 0.5" ($12.7 \unit{\milli \meter}$) tooling ball. By measuring the position of set of 0.5" tooling balls in the panel with the camera array in a station-wide coordinate system, we map the X-ray measurements to a station-wide reference frame.

\subsection{Tracker Panel Fiducials}
\label{sect:fids}

In this photogrammetry technique, we used the Carr Lane CL-10-SCB tooling balls with a pin diameter tolerance of .2500" ($6.350 \unit{\milli \meter}$)/.2496 ($6.3398 \unit{\milli \meter}$) and a ball diameter tolerance of .5002/.49998. Inserting the tooling balls into the panel for an extended period of time could deflate the straw tubes if the room pressure increased sufficiently. This could cause damage to the straw tubes. To allow air flow, we drilled a small hole in each of the tooling balls along the tooling ball axis. We avoided any distortion of the image analysis by just excluding the pixels near the top edge of the ball from the fit. 

Unfortunately, we found that many of the panels contained significant epoxy on many of the fiducial spot faces; this is thought to be from panel production. This epoxy was discovered $\it{after}$ the panel X-ray scan. Therefore, in the X-ray survey the fiducial may be resting on epoxy thus positioned radially outward and may be at an angle. 

\begin{figure}[htb]
        \centering
        \includegraphics[width=0.2\textwidth]{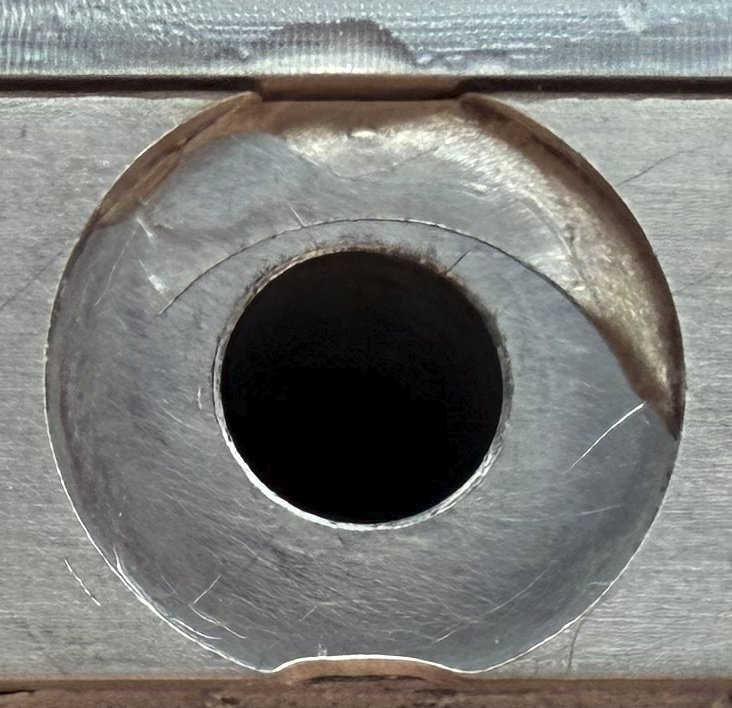}
        \caption{Epoxy on the fiducial spot face}
    \label{Epoxy}
\end{figure}

\begin{figure*}[htb]
    \centering
    \begin{subfigure}[t]{0.21\textwidth}
        \centering
        \includegraphics[width=0.95\textwidth]{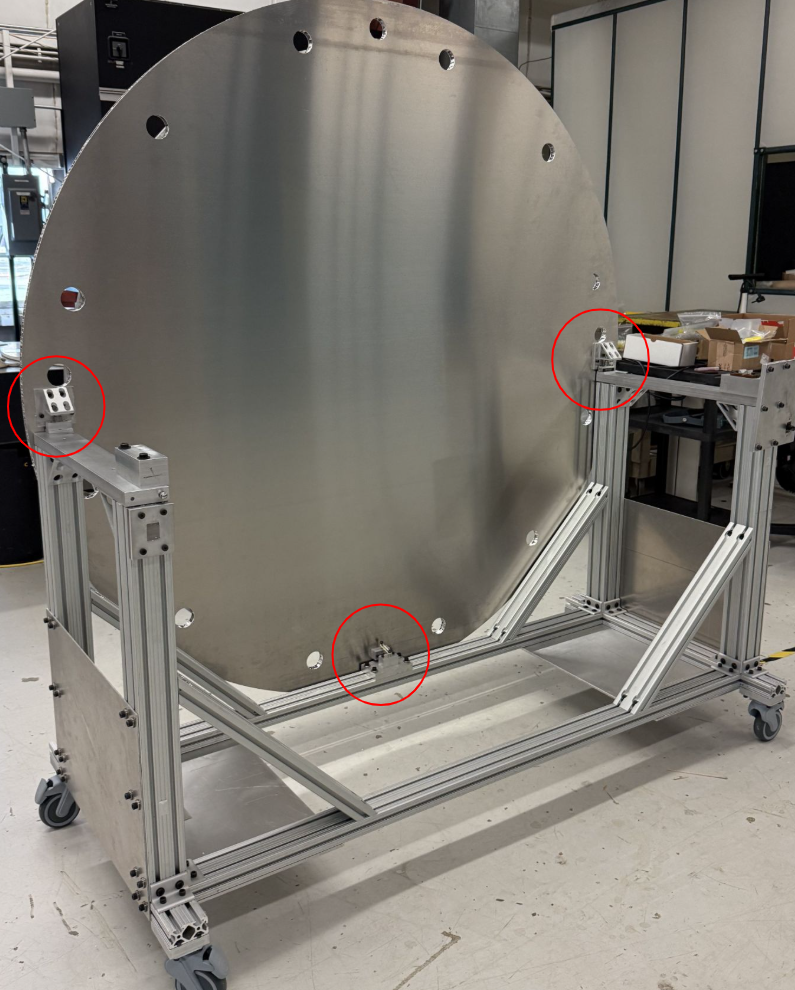}
        \caption{}
    \end{subfigure}
    \begin{subfigure}[t]{0.22\textwidth}
        \centering
        \includegraphics[width=0.95\textwidth]{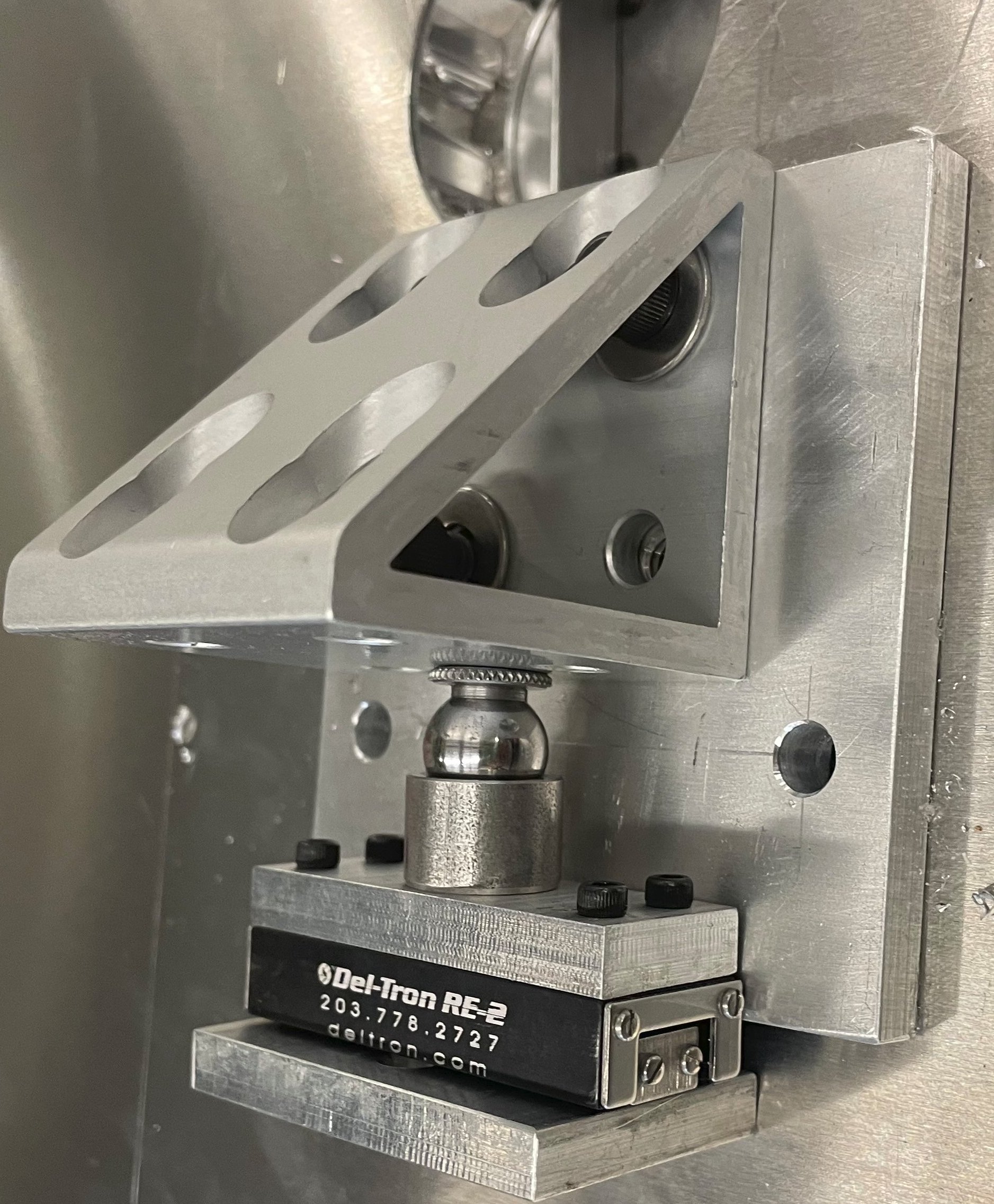}
        \caption{}
    \end{subfigure}
    \begin{subfigure}[t]{0.2\textwidth}
        \centering
        \includegraphics[width=0.95\textwidth]{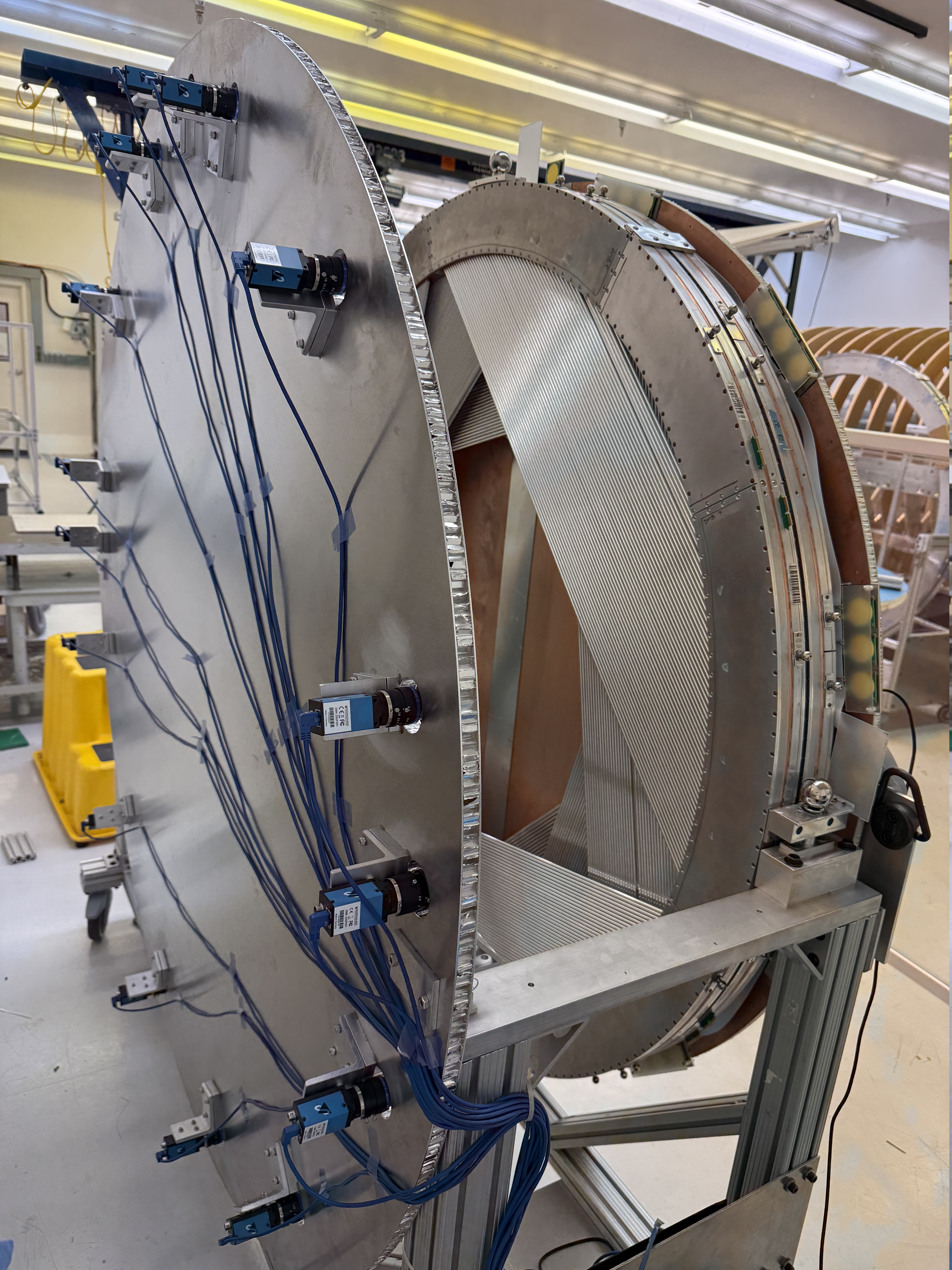}
        \caption{}
    \end{subfigure}
        \begin{subfigure}[t]{0.18\textwidth}
        \centering
        \includegraphics[width=0.95\textwidth]{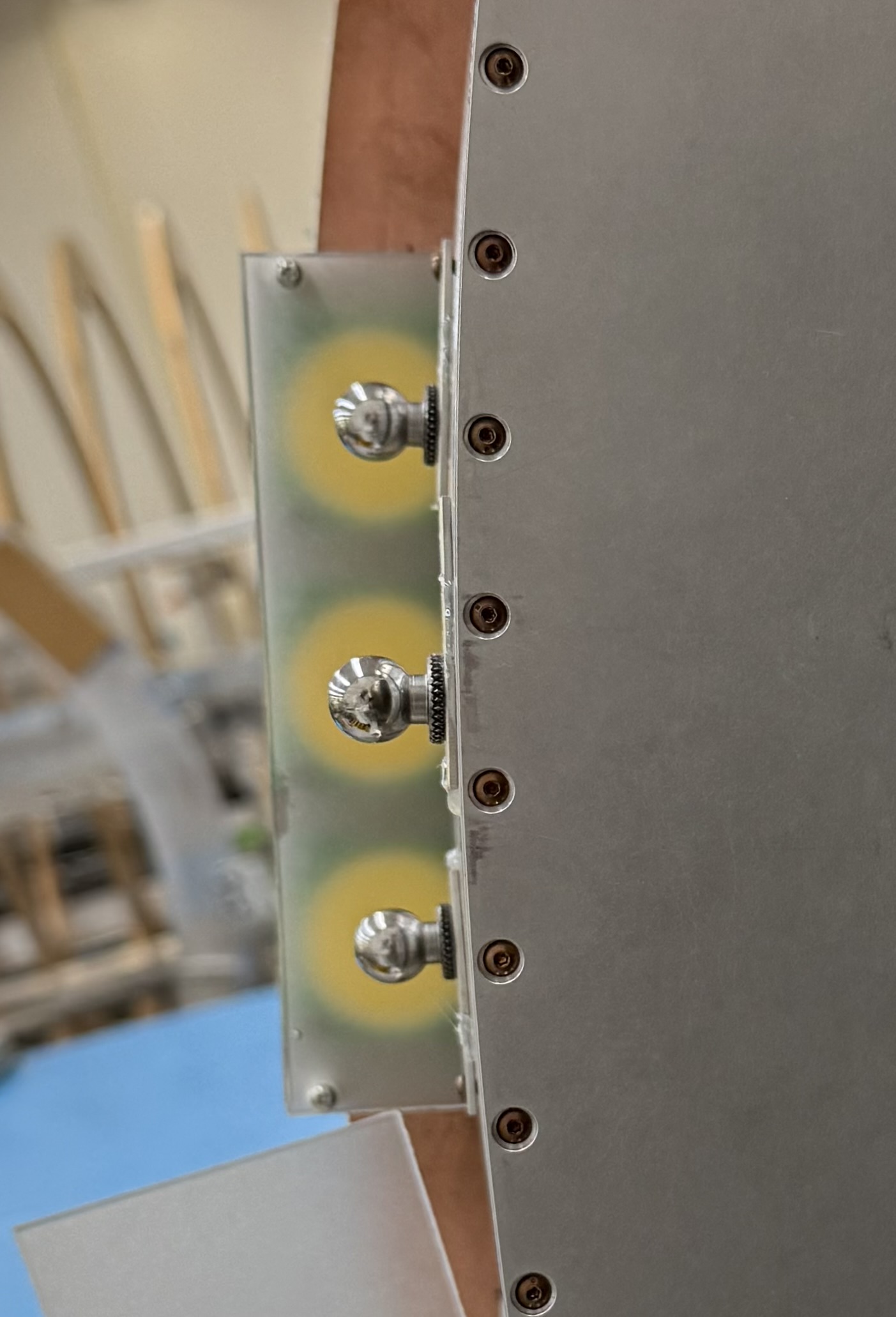}
        \caption{}
    \end{subfigure}
    \caption{Figures detail features of the metrology stand. (a) Metrology stand without station. The kinematic mounts are circled in red. (b) 1 of 3 kinematic mounts mounting  honeycomb on the 80-20 stand. (c) Station inside the metrology stand detailing the camera mounts on the Aluminum honeycomb. (d) Tooling balls on station and LEDs viewed along the camera axis.}
    \label{metrologystand}
\end{figure*}

In order for the measurements to be consistent with the X-ray scan, the epoxy was kept intact. The epoxy is not infinitely rigid, so the radial position of the fiducial is dependent on how much
pressure is applied on the epoxy surface. In addition, in some cases this epoxy was accidentally removed in between the X-ray scan and the camera measurement; this will change the length between the fiducials. An image of the epoxy on the spot faces is shown in Figure \ref{Epoxy}. We will discuss this effect in the results section.

\subsection{Camera Metrology Stand}
\label{sect:stand}

The metrology stand is shown in Figure \ref{metrologystand}. The stand is made of $80/20$\textregistered \hspace{0.1mm} aluminum extrusions with aluminum bars and plates. 

There are a total of 15 cameras on the metrology stand, which are mounted on a single Aluminum honeycomb disk. The cameras and lenses are all from The Imaging Source. The cameras are $\sim 5$M pixels with CCD dimension of 2592 x 1944 with a pixel size of  $2.2 \unit{\micro\meter}$ (DMK 33GP031). Twelve of the cameras view triplets of tooling balls on the panels and are equipped with $25 \unit{\milli \meter}$ focal length lenses (TCL 2518 5MP). The final three cameras view single 1.5" ($38.1 \unit{\milli \meter}$) fiducials; to take advantage of the full field of view, the cameras use a $50 \unit{\milli \meter}$ lens (TCL 5026 5MP). Each camera is first individually mounted on an angle bracket. The bracket is attached to a transition plate, which is epoxied onto the honeycomb. The camera focus and aperture were fixed throughout station production. To optimize the two parameters, each camera was individually placed on a bench-top setup with a triplet of tooling balls placed the appropriate distance away to match that in the camera stand. First the aperture was opened wide to narrow the depth of field, then the focus was optimized and fixed. Finally the aperture was set near the minimum to maximize the depth of field. 

Note, in early testing with a single camera on a translation stage we observed significant improvements to the radius measurement with this lens over less expensive options. We did not test less expensive cameras other than an inexpensive web-cam, which was not sufficient. In general, the 5 MP camera and the improved lens were critical to achieve a precision measurement of the tooling ball radius and therefore the position along the camera axis.

The honeycomb sheet is a $\sim 6$ ft diameter disk consisting of two outer layers of 0.063" ($1.6002 \unit{\milli \meter}$) thick aluminum and a 3/8" ($9.525 \unit{\milli \meter}$) inner aluminum honeycomb. We selected the honeycomb as it offers rigidity while also being lightweight. The honeycomb was purchased from Pacific Panels. The honeycomb is mounted on the stand using another three point kinematic mount: ball/cone connection defining XYZ, ball/cone on a sliding stage defining YZ, and a ball against a flat plate defining Z. This kinematic mount is to avoid any deformations of the honeycomb. 

Note, to maintain high-precision measurements, it is critical to avoid deformations of the honeycomb. Any local deformation of the honeycomb of $\sim 0.01 \unit{\milli \meter}$ across the span of the camera width would result in a $0.15 \unit{\milli \meter}$ shift in the apparent tooling ball position. The honeycomb is mounted just below the stand's vertical extrusions to avoid any stand deformations. Checks of this are discussed in Section \ref{sect:consistency}.

On the other side of the stand, the station is mounted in the same three point kinematic mount used in the tracker frame. To handle the even and odd stations, the stations are moved into the metrology stand with the same orientation; i.e. for odd stations, the stand kinematic mounts are swapped (kinematic V-groove and cone). This is relevant as this switches which side of the stand defines the X position of the stations. 
\begin{figure*}[hb]
        \centering
        \includegraphics[width=0.7\textwidth]{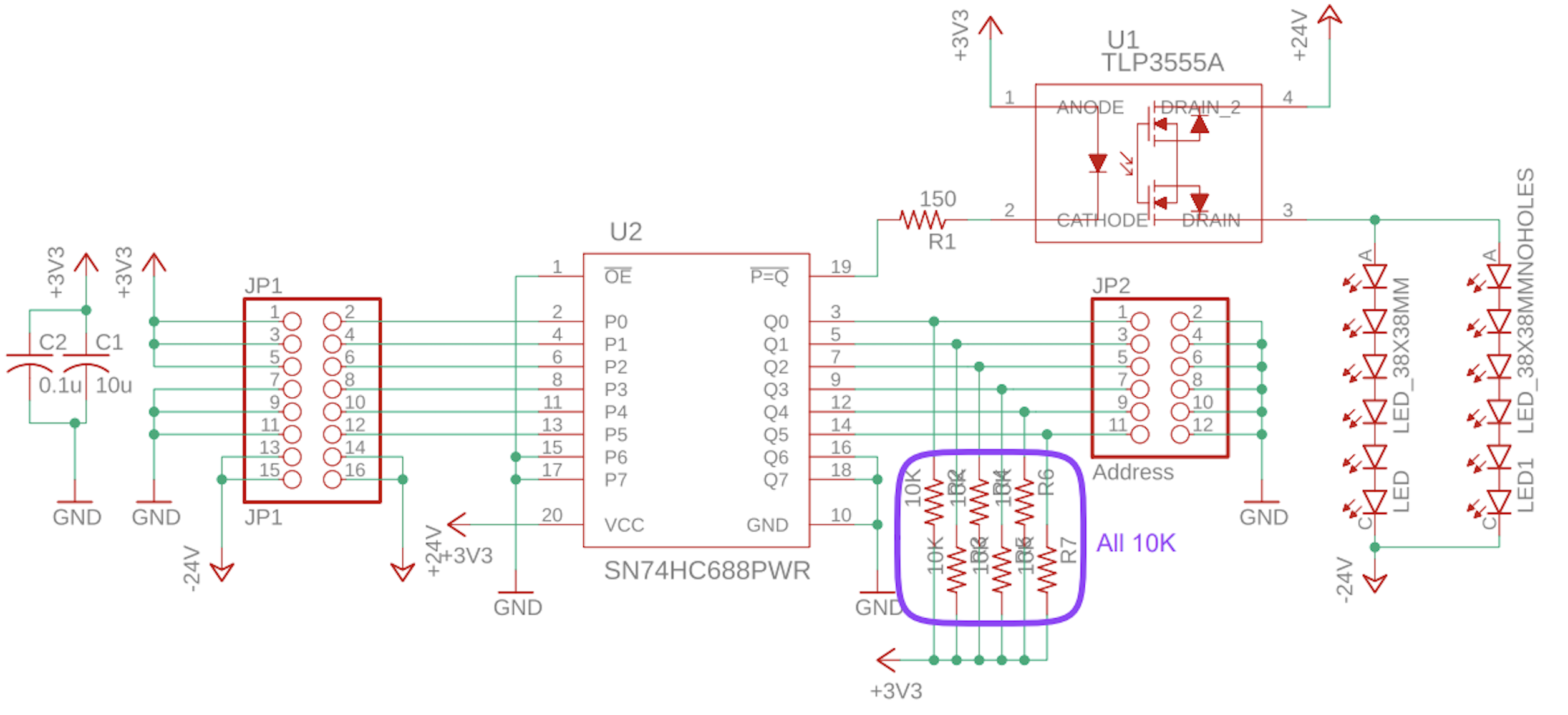}
        \caption{LED Schematic Diagram}
    \label{LED}
\end{figure*}
The last component of the stand is the LED array, which is used as a back-light. In preliminary camera tests, we found that the position and radius of the tooling balls was very sensitive to background conditions. Even with a white background, we found systematics due to room lighting. For example with room lights on, the top of the ball would be brighter resulting in pixel-by-pixel residuals associated with a larger apparent radius. To avoid this, we used an LED as a background for each individual fiducial.

A single LED board diagram is shown in Figure \ref{LED}. The LED itself is from Luminus Devices Inc. (CXM-32-50-80-54-AC30-F4-3) offering 162 lm/W. The high lumen LEDs allow for a very short exposure time making the background lighting negligible.  The LEDs are significantly larger than the tooling ball such that the LED provides uniform lighting even with O($\unit{\milli \meter}$) shifts in the tooling ball position with respect to the LED (e.g. due to panel-panel differences in the station construction). The LEDs are connected in series each with a unique serial address (0-38). Both the LEDs and the cameras were controlled by the same computer; each camera captures one image per fiducial.

\subsection{Operations and Workflow}
\label{sect:op}
After station assembly, the metrology was initiated by the mechanical measurements which took $<$ 1 hour per station. Once the station was mounted in the metrology stand, a single python script communicated with all 15 cameras and 39 LEDs to take 4 images per LED (156 images). The image analysis and processing of the images took $< 5$ minutes resulting in the position and radius of the tooling balls on the camera CCDs. 

All of the data was uploaded to a database for additional handling. This included the fitted tooling ball positions and radii, the station panel IDs, the mechanical measurements, etc. In addition the X-ray surveyed data, and reference positions on the tracker frame were uploaded to this database. 

A python script pulled the data, and then transformed the X-ray surveyed data into the station-wide coordinate system and the frame coordinate system. This took $< 1$ minute per station.

\subsection{Coordinate Systems}
\label{sect:coordinateSystem}

 Here, we describe the relevant coordinate systems to the metrology. As an overview, there is the tracker-wide coordinate system, and a coordinate system in the metrology stand. Each camera has a 3D coordinate system defined by the position and orientation of the camera as well as a 2D coordinate system on each camera's CCD.
 
In the Mu2e coordinate system, $\vec{X}_{Mu2e}$, $Z_{Mu2e}$ is along the beam axis, $Y_{Mu2e}$ is vertical, and $X_{Mu2e}$ is defined such that the coordinate system is a right-handed coordinate system. 
 
\begin{figure}[hbt]
{\centering
\includegraphics[width=0.4\textwidth]{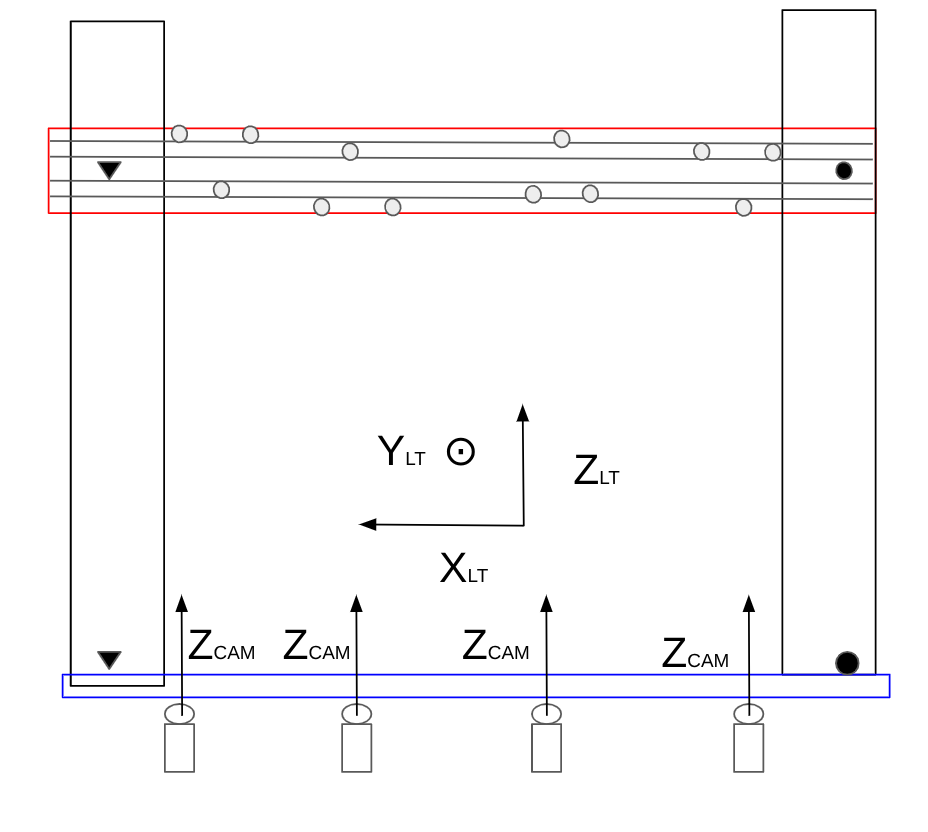}
\caption{The relevant coordinate systems in the camera stand. The station is shown in red, the honeycomb with the mounted cameras is shown in blue. Each camera has its own coordinate system. Example tooling ball triplets are shown on the station. The dark triangles and balls represent the station's and the honeycomb's kinematic mounts. 
\label{CoordinateSystem}
}}\end{figure}

 The additional coordinate systems associated with the metrology stand are shown in Figure \ref{CoordinateSystem}. The station-wide coordinate system measured by the laser tracker is defined as $\vec{X}_{LT}$. 

 This coordinate system has the same XYZ convention as $\vec{X}_{Mu2e}$, but has the origin centered on the calibration station (Section \ref{sect:calibrationstation}). The direction $Y_{LT}$ is defined by gravity, and the direction of $X_{LT}/Z_{LT}$ is defined using fiducials on the stand (one at $+X_{LT}$ and one at $-X_{LT}$).  

 In this coordinate system, the laser tracker measured the reference fiducials on the calibration station, the three kinematic mount reference features: two 1/8" ($3.175 \unit{\milli \meter}$) pin positions and the bottom slot), and a few reference fiducials on the stand itself. These kinematic mount reference features and the stand fiducials were measured before the calibration station was inserted into the stand.

 In addition, each camera has its own reference frame, $\vec{X}_{CAM}$. The camera frame origin is defined as the center of the camera's effective lens and $Z_{CAM}$ is along the camera axis. The cameras are mounted $\sim 0.5$ m away from the station in $Z_{CAM}$ such that the $X_{CAM}/Y_{CAM}$ origin is centered about the central tooling ball in each triplet of tooling balls.  Each camera's $Z_{CAM}$ axis is nominally aligned with $Z_{LT}$, but each camera is rotated about the $Z_{LT}$ axis in $\sim 30$ degree increments such that the triplet of tooling balls on the radial extent of the station form a line in ${X}_{CAM}$. Finally, the CCD frame is a 2D frame on each camera's image plane aligned with the camera's CCD; $X_{CCD}/Y_{CCD}$ are parallel to $X_{CAM}/Y_{CAM}$.

\subsection{Calibration Station}
\label{sect:calibrationstation}

\begin{figure}[htb]
    \centering
        \includegraphics[angle=270, width=0.3\textwidth]{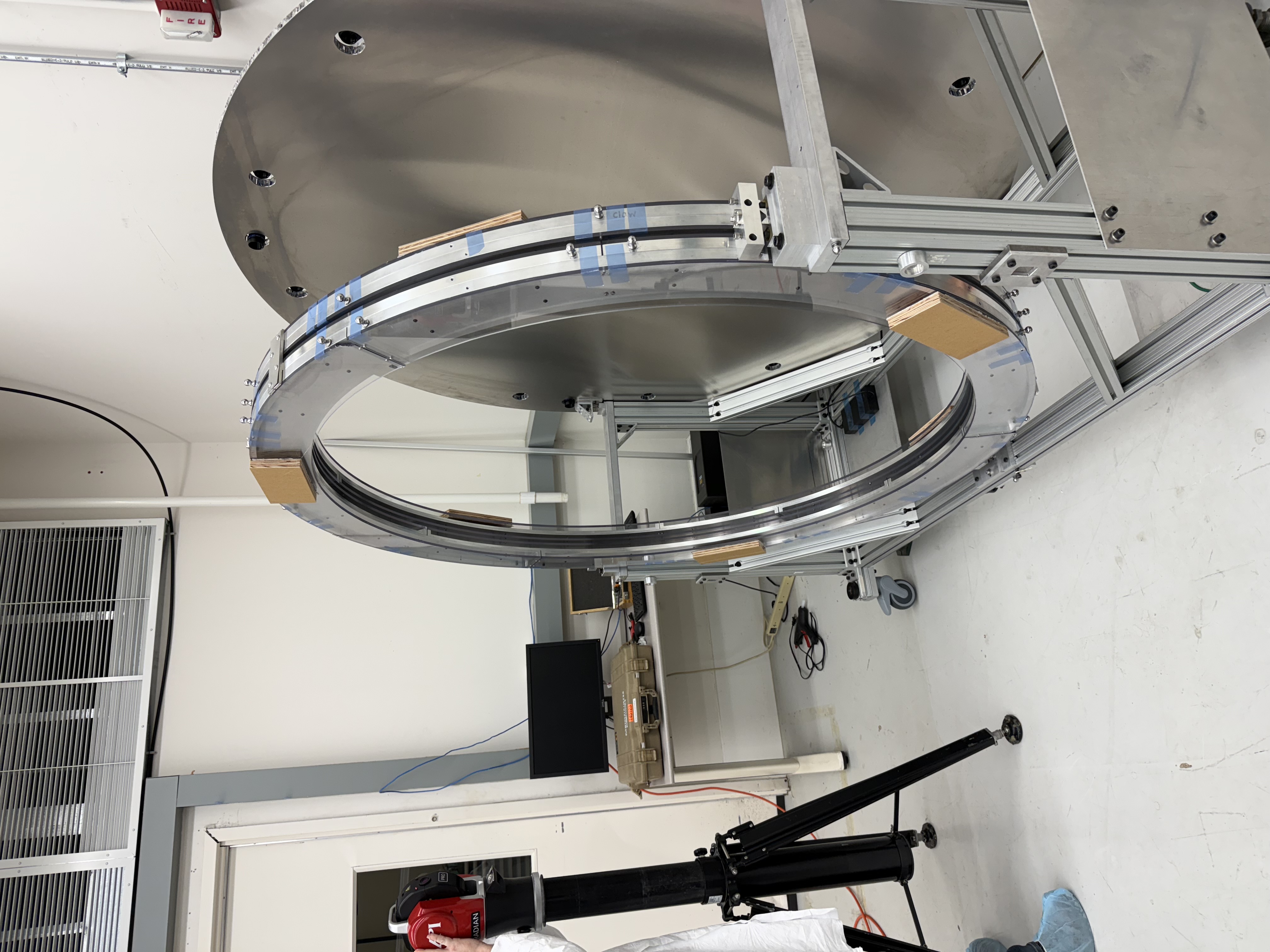}
    \caption{The calibration station in the metrology stand during the laser tracker survey. }
    \label{calibrationstation}
\end{figure}

The calibration station is the metrological reference artifact
 for the camera array; the station in the metrology stand is shown in Figure \ref{calibrationstation}. Note, the station also is used to verify the stability of the camera array in between production stations. The calibration station contains all the fiducials of a production station. These fiducials were screwed and epoxied in place to avoid any movement over time. The panels in the calibration station were also epoxied together to avoid any deformations of the station. 

The fiducials were surveyed by a laser tracker. The laser tracker measured O(15) points per reference fiducial and fit for the center of each tooling ball.  Based on tests with the laser tracker over a similar span, we expect a resolution of $\sim 0.05 \si{\milli\meter}$

\subsection{Image Analysis}
\label{sect:ImageAnalysis}
\begin{figure}[hbt]
{\centering
\includegraphics[width=0.4\textwidth]{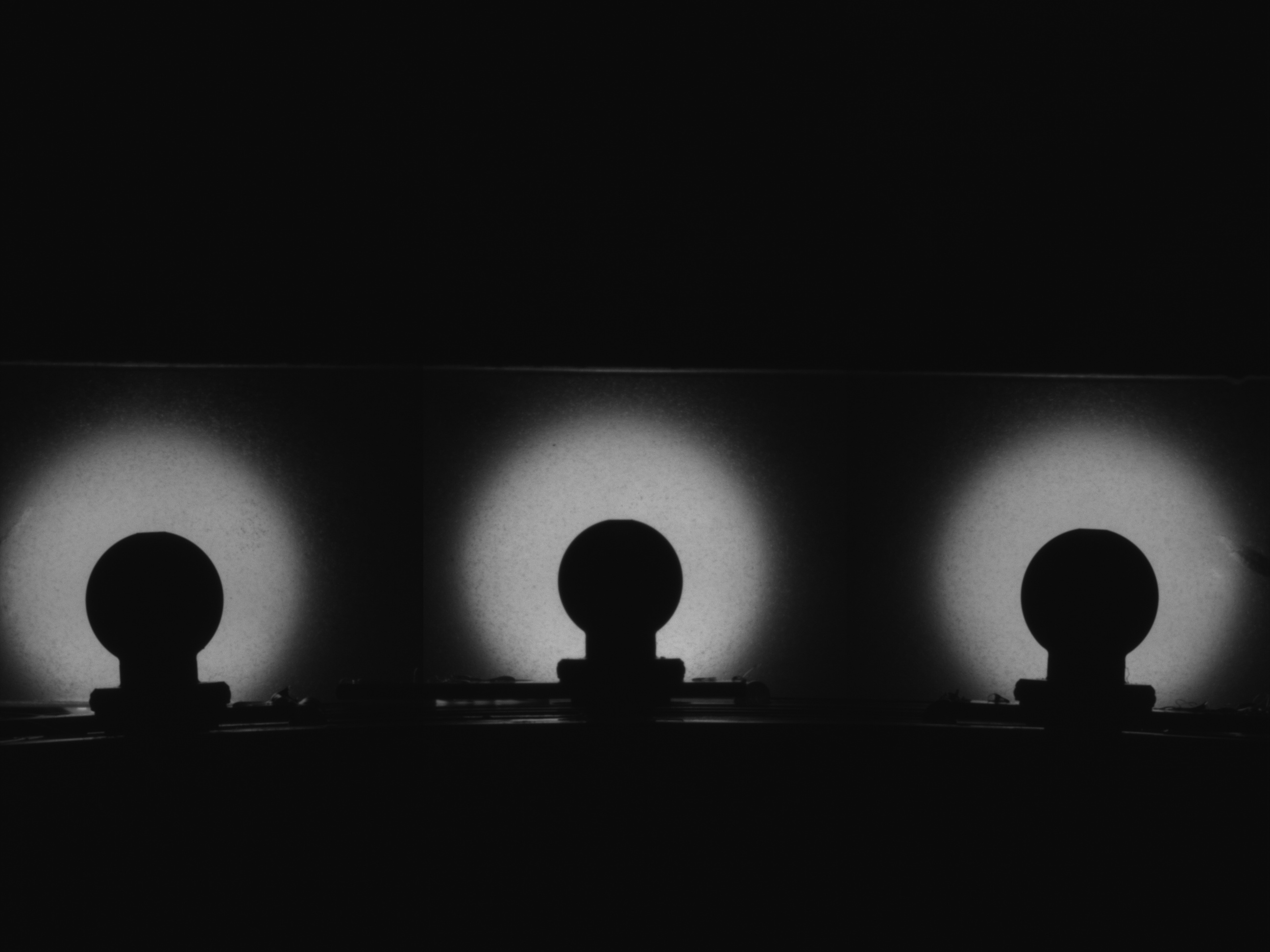}
\caption{A triplet of tooling balls imaged by a camera. Note, these 3 tooling balls are from different panels. }
\label{Triplet}
}\end{figure}

The objective of the image analysis is to identify pixels associated with the edge of each tooling ball and fit the 2D pixel coordinates to estimate the 2D center and radius of the tooling ball in $X_{CCD}$. An image of a triplet of tooling balls taken by a camera is shown in Figure \ref{Triplet}. This is actually the superposition of three separate images where each LED was only turned on once. Turning the LEDs on simultaneously results in systematic residuals in the tooling ball radius measurement described below. 

The analysis is initiated by searching for the tooling balls on the image plane ($X_{CCD}$). This is first done using a Hough circle algorithm\cite{duda1972use}; this yields a starting position of each tooling ball. We found the radius of the tooling ball is particularly sensitive to the LED lighting intensity. 

\begin{figure}[hbt]
{\centering
\includegraphics[width=0.5\textwidth]{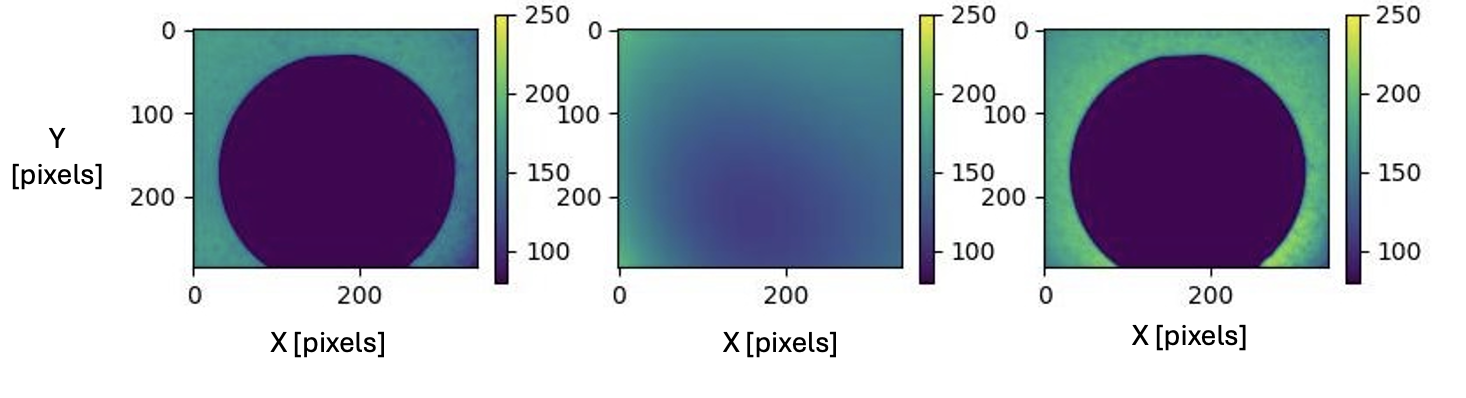}
\caption{A zoom-in on an individual tooling ball showing variation in the background lighting. The second image shows the fitted background lighting. The third image shows tooling ball with the corrected background lighting. Note all images are scaled to the same average background intensity. 
\label{BkgLight}
}}\end{figure}

To eliminate this systematic in each image we fit for a 2D background intensity (polynomial fit) around the Hough circle positions and then apply a correction to achieve uniform lighting around the tooling ball.
 Figure \ref{BkgLight} shows the original image, the fitted background intensity, and the image with the corrected background intensity. 
 
\begin{figure}[htb]
{\centering
\includegraphics[width=0.5\textwidth]{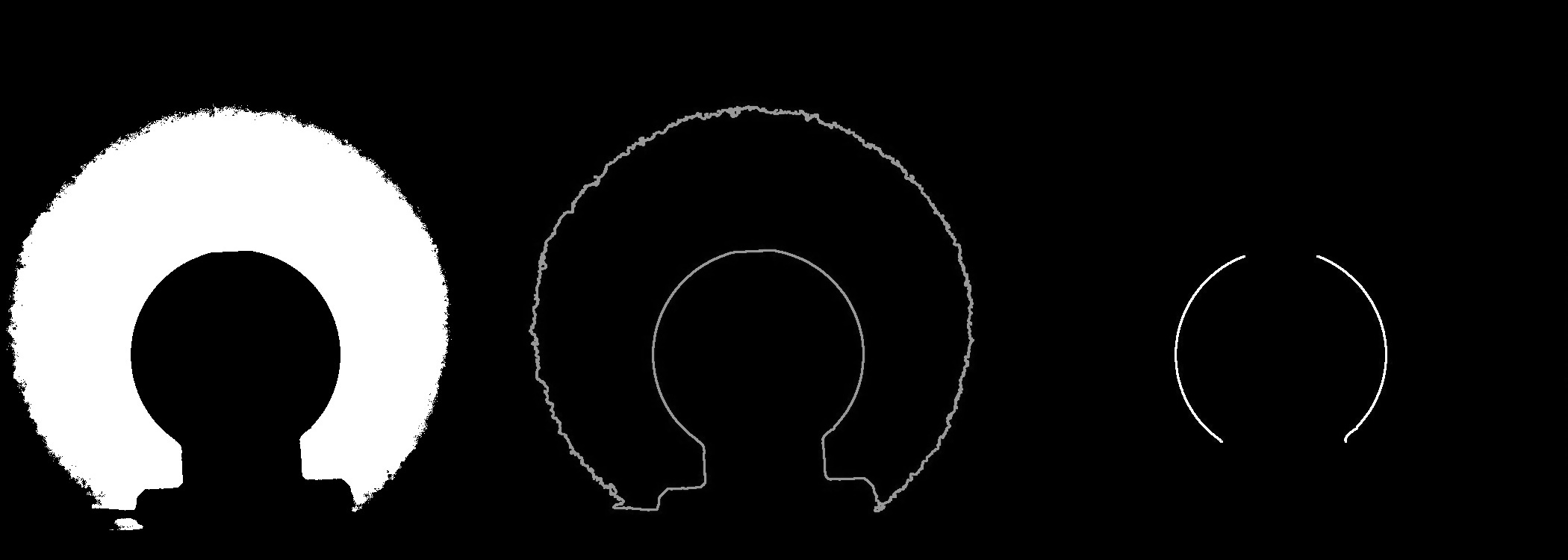}
\caption{Image analysis stages. The first subplot shows a binary image, the second shows the relevant contours, the third is the filtered contour. 
\label{ImageAnalysisComb}
 }}\end{figure}

We apply a binary threshold to the corrected image. This is done in image analysis software package OpenCV\cite{opencv_library}. We select all pixels at the threshold over the full image. We then filter the pixels to select only those near the three tooling balls found by the Hough circle algorithm. In addition, we apply a cut to remove pixels near the stem of the tooling ball and the top of the tooling ball to avoid the hole used to allow air to flow in/out of the panels. Figure \ref{ImageAnalysisComb} shows stages of the image analysis for the left-most tooling ball.

\begin{figure}[htb]
{\centering
\includegraphics[width=0.4\textwidth]{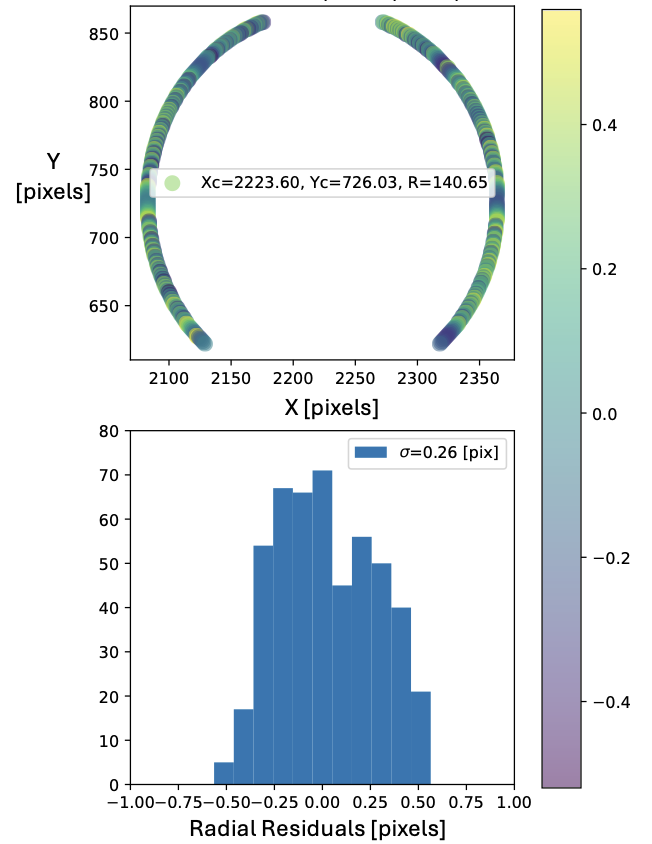}
\caption{The first row represents the radial residuals as a function of XY pixel. The clustering of high or low residuals represents some non-uniformity like dust on the tooling ball, background light variations, etc. The second row is a histogram of the radial residuals for a single tooling ball fit. 
\label{ImageResiduals}
}}\end{figure}

These filtered pixels on the threshold contour are fit for a 2D position and radius of the tooling ball on the camera's CCD $X_{CCD}, Y_{CCD}, R_{CCD}$. The radial residuals as a function of position on the CCD and a histogram of the radial residuals are shown in Figure \ref{ImageResiduals}. We average the image analysis over four images to yield an improved measurement.

\subsection{Local Mechanical Measurements}
\label{sect:camAxis}
\begin{figure*}[ht]
    \centering
    \begin{subfigure}[t]{0.25\textwidth}
        \centering
        \includegraphics[width=0.9\textwidth]{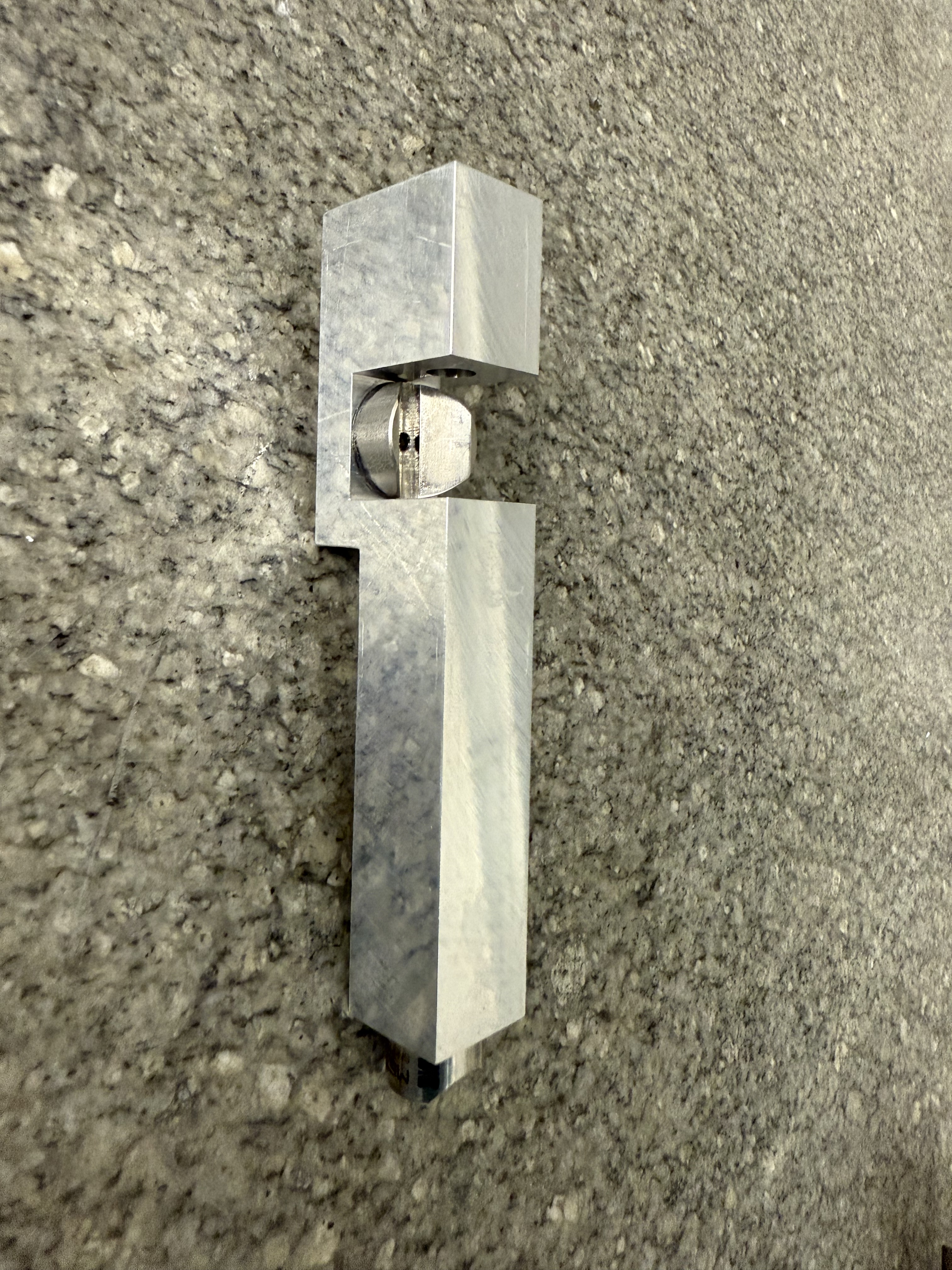}
        \caption{}
    \end{subfigure}
    \begin{subfigure}[t]{0.25\textwidth}
        \centering
        \includegraphics[width=0.9\textwidth]{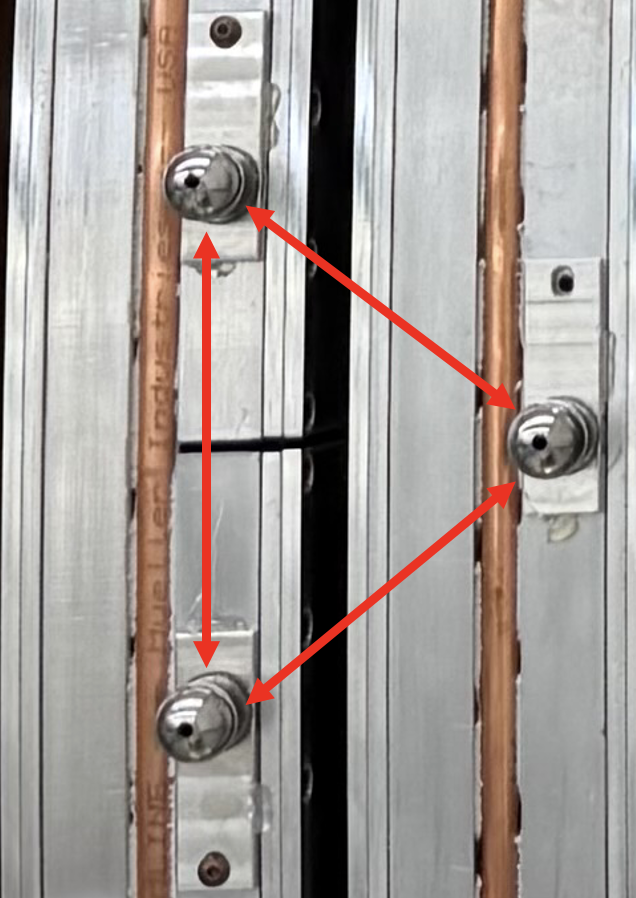}
        \caption{}
    \end{subfigure}
    \begin{subfigure}[t]{0.4\textwidth}
        \centering
        \includegraphics[width=0.9\textwidth]{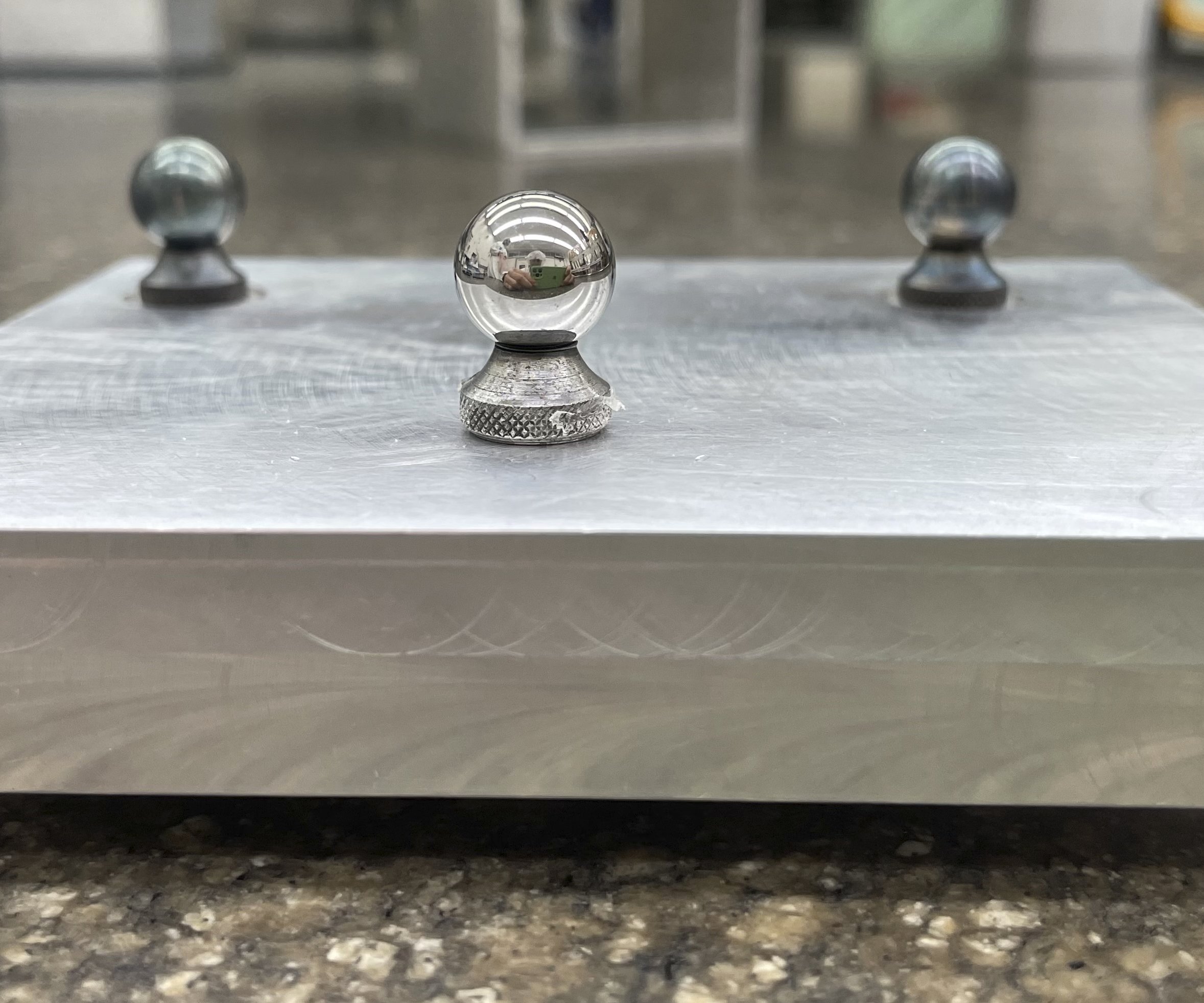}
        \caption{}
        \label{Calibrationblock}
    \end{subfigure}
    
    \caption{The figure details the mechanical measurement jig and its calibration block. (a) Mechanical measurement jig: cone engages 1 ball, V-groove engages another.  (b) Lengths measured mechanically. Note, all three fiducials are on different panels.  The longest length is between two panels on the same plane. (c) Calibration block used to calibrate the mechanical measurement dial indicator.}
    \label{MechanicalJig}
\end{figure*}
Using optical projection equations, the apparent fiducial radius on the CCD is used to reconstruct the position along the camera axis. 

We are using 0.5" ($6.35 \unit{\milli \meter}$ radius)  tooling balls roughly $500 \unit{\milli \meter}$ away from the camera. To achieve $0.1 \unit{\milli \meter}$ over the $500 \unit{\milli \meter}$ distance, one requires a fractional precision of $2\cdot 10^{-4}$; this corresponds to a radius measurement of $6.35 \cdot 2\cdot 10^{-4}=1.3 \unit{\micro\meter}$. Even with the background LEDs, averaging over several images, this is extremely difficult and will be sensitive to temperature differences and systematics in the radius reconstruction. 

Therefore, we decided to make additional measurements to achieve a position resolution of $100 \unit{\micro\meter}$ along the camera axis ($Z_{CAM}$). The limitation of using the tooling ball radius to measure the distance from the camera ($Z_{CAM}$) is the size of the tooling ball ($6.35 \unit{\milli \meter}$ radius used to measure a distance of $\sim 500 \unit{\milli \meter}$). By instead measuring the distance between the tooling balls (longest length of $86.1 \unit{\milli \meter}$), we can achieve a significantly better measurement along the camera axis. We created two jigs (Figure \ref{MechanicalJig}) to mechanically measure the three lengths between the fiducials in each triplet of fiducials.

The jigs contain a kinematic cone to define the XYZ position of one tooling ball and a kinematic V groove to define the YZ position of the second tooling ball. The local X position of the second tooling ball with respect to the first tooling ball is measured using a dial indicator with 0.0001" precision ($2.5 \unit{\micro\meter}$). 

The image analysis then uses the 2D position of the tooling balls on the camera's CCD and the 3 mechanical measurements.  The stability of these measurements is discussed in Section \ref{sec:mechresults}.

By adjusting the longest of the three dial indicator measurements in the image analysis by $\pm \epsilon$, we observe that all three tooling balls move in $Z_{CAM}$. This measurement effectively sets the $Z_{CAM}$ for the three fiducials. The other two measurements define the relative $Z_{CAM}$ position of the three fiducials. Therefore, we are particularly sensitive to the longest dial indicator measurement. A $25 \unit{\micro\meter}$ measurement error results in a $Z_{CAM}$ error of $150 \unit{\micro\meter}$ for all three fiducials. 

The dial indicator calibration is completed using a triplet of tooling balls with a known distance between them. This calibration block (Figure \ref{Calibrationblock}) was machined to match the average distances expected in a triplet of tooling balls.

\subsection{Optical Projection}
\label{sect:opticalProject}
In this section, we describe the optical projection using the tooling ball 2D position along with the radius and discuss including the local dial-indicator length measurements between fiducials.  

We use the following optical equations, where f is the camera's focal length and $d_{I}$ is the distance from the lens to the image plane. The focal length was optimized by minimizing the difference between the lengths between the fiducials measured by the laser tracker and that using the measured tooling ball radius for the 12 cameras (Section \ref{sect:TransformationCameraLaserTracker}). The focal length was optimized to be $26.5 \unit{\milli \meter}$ compared to $25 \unit{\milli \meter}$ listed from the manufacturer. 
  
$$\frac{1}{f} = \frac{1}{d_{I}} + \frac{1}{Z_{CAM}}, \frac{d_{I}}{Z_{CAM}} = \frac{X_{CCD}}{X_{CAM}}, \frac{d_{I}}{Z_{CAM}} = \frac{Y_{CCD}}{Y_{CAM}}$$

Eliminating the distance to the image plane, we are left with the following projection equations. 
$$R_{CCD} = \frac{R_{CAM} \cdot f} {Z_{CAM} - f}$$ 

$$X_{CCD} = \frac{X_{CAM} \cdot f} {Z_{CAM} - f}, Y_{CCD} = \frac{Y_{CAM} \cdot f} {Z_{CAM} - f}$$

By combining the fitted radius on the camera's CCD with the known radius from the tooling ball manufacturer, we determine $Z_{CAM}$ for each tooling ball. Inserting $Z_{CAM}$ and $X_{CCD},Y_{CCD}$ into the above equations determines $X_{CAM}, Y_{CAM}$. 

Our final 3D measurement in the camera frame is the result of the 2D position of each tooling ball on the camera CCD (2*3 equations) plus the three length constraints from the mechanical measurements.

This is a combined 9 equations we can fit for the 9 unknowns. We solve the system of equations numerically using Sympy\cite{sympy}. Note, Sympy finds several solutions, but only one physical solution for each fit: $Z_{CAM}>0$ for all tooling balls, the central tooling ball is either closer/farther to the camera than the side tooling balls.

\subsection{Transformation From Camera to Station-Wide Frame}
\label{sect:TransformationCameraLaserTracker}

A laser tracker survey determines the position of the calibration station fiducials with respect to the metrology stand ($\vec{X}_{LT}$). Employing the above image analysis to images of the calibration station fiducials yields the position of the calibration station fiducials in the camera frame ($\vec{X}_{CAM}$).

We fit for the transformation from $\vec{X}_{CAM}$ to $\vec{X}_{LT}$. We use a ZYZ Euler angle rotation convention plus a 3D translation. We apply the rotation to the triplet of balls (not the reference frame).    

That is: 
$$\vec{X}_{CAM}^{*} = Z(Y(Z(\vec{X}_{CAM}, \phi), \theta), \psi) + \vec{dX}_{CAM}$$

We minimize the sum of squares listed below:

\begin{equation}
\label{CamChi2}
\begin{aligned}
SS^{2} = \sum_{i,CAM}^{N}  (X_{i, CAM}^{*} - X_{i, LT})^{2} + \\
(Y_{i, CAM}^{*} - Y_{i, LT})^{2} + \\
(Z_{i, CAM}^{*} - Z_{i, LT})^{2}
\end{aligned}
\end{equation}

This minimization sums over the three fiducials in a triplet of tooling balls viewed by the camera. We minimize using the python Scipy\cite{scipy} package using the Nelder-Mead minimization routine\cite{nelder}. Here, we are applying the above mentioned rotations/translations to the camera frame to match the laser tracker coordinates. For all production stations, the data in the camera frame can be transformed into the station-wide laser tracker frame using these transformations.

\subsection{Transforming Panel Data to the Laser Tracker Frame}
\label{sect:TransformationPanelToLaserTracker}

For each production station, after transforming the image data into the station-wide laser tracker frame, we fit for the transformation from the individual panel frames to the laser tracker frame for each of the 12 panels/station. This is transforming the X-ray data for each panel into the station-wide coordinate system. We use the same ZYZ Euler angle transformation described above and a similar sum of squares: 

\begin{equation}
\label{PanelChi2}
\begin{aligned}
SS = \sum_{i,P}^{N} 
(X_{i, P}^{*} - X_{i, LT})^{2} + \\
(Y_{i, P}^{*} - Y_{i, LT})^{2} + \\
(Z_{i, P}^{*} - Z_{i, LT})^{2}
\end{aligned}
\end{equation}

This is summing over the three fiducials in a $\it{panel}$, which is the result of the image analysis from three separate cameras. 

There are three fiducials per panel, each with an XYZ position (9 total DOF). We fit for the 6 parameter rigid-body transformation from the local X-ray panel coordinate system to the panel position in the station-wide laser tracker coordinate system. There are 3 remaining degrees of freedom, which correspond to the lengths between the three fiducials, roughly 784, 784 and 1379 mm.

\subsection{Tracker Frame CMM Measurement and Transforming Photogrammetry Data into the Frame}
\label{sect:TransformationStationWideToFrame}
In this section we describe the procedure to transform the station fiducials measured in the metrology stand into a tracker-wide coordinate system. 

Note, before the stations were inserted into the tracker frame, the frame itself was sent to Fermilab's large Coordinate Measuring Machine (CMM). The CMM measured the 3D position of the 1/8" ($3.175 \unit{\milli \meter}$) pin holes and the slots at the bottom of the frame (shown in Figure \ref{framefids}) to high precision (<0.001" or 25 $\unit{\micro \meter}$). Those references were measured with and without weight to simulate the effect of loading the stations into the frame. The weight changes the position by O($100 \unit{\micro\meter}$). 

The position of the kinematic mounts in the metrology stand and the tracker frame are defined with respect to these 1/8" ($3.175 \unit{\milli \meter}$) pin holes. The kinematic mounts fully define the position and orientation of the stations in both the metrology stand and the tracker frame. Therefore assuming the same kinematic mount, no deformations of the stations and no deformations of the tracker frame since the CMM, we can determine the rigid body transformation to transform data in the metrology stand coordinate system to the tracker frame for a given slot.

We are assuming that the relative panel to panel alignment does not change as we move the stations into the frame. Tests with the first two stations suggested that the panel-panel alignment within a station is not changing at the $100 \unit{\micro\meter}$ level even if a station is put horizontal again on a table and then back into the metrology stand. 

In addition, there is the possibility that the frame will distort and therefore distort the station-to-station alignment. For example, one 'stave' (defining the position of the stations at +X) can shift in $Z_{Mu2e}$ with respect to the other (defining the position of the stations at -X). Features on the tracker frame exterior were re-measured once the fully-assembled tracker was positioned in the Mu2e experimental hall, therefore capturing any change in the frame distortion. 

 With this, we produce a modified position of the frame references and therefore a modified position of the stations in the frame. 

\subsection{Station-To-Station Alignment Optimization and Validation}
\label{StationOptimization}
Using the frame CMM, measurements of the tracker frame in the hall, and the station metrology we determine the position of the fiducials and the straw tubes in the full tracker assuming the \textit{same} kinematic mount used in the metrology stand. Note, the stations were transported to the hall on transport blocks, not their kinematic mounts. 

The transformed positions of the fiducials in the frame indicate station-to-station misalignments with respect to nominal. This is partially due to the shape of the frame with respect to nominal and partially due to station-to-station variation in construction. We correct for these station-to-station misalignments by custom-machining the kinematic mounts for each station. For both the cone-ball (XYZ) and the canoe sphere-V (YZ) connections, the cone or V land in a 3/4" ($19.05 \unit{\milli \meter}$) counter-bore. By adjusting the counter-bore's depth and position with respect to the stave pin hole, we optimize the position and rotation of the individual stations with respect to nominal. This requires machining a precision position, depth, and radius of the counter-bores with respect to the pin hole ideally to $\sim 25 \unit{\micro \meter}$.

There is also an additional measurement of the station-to-station alignment built into the metrology procedure once all the stations are positioned in their custom kinematic mounts. A laser tracker can measure the three station-wide fiducials per station. Since these were also measured in the camera metrology stand for each station, this can serve as an independent transformation of the metrology data from the metrology stand to the full tracker frame.  Compared to the panel fiducials, these station fiducials are easier to access and less disruptive to tracker activities. This measurement will likely be done sometime in the future. This also can be used as a validation of the custom mounts.

\section{Results}
\label{sect:Results}
In this section we describe the results from the relevant transformations and tests to determine the precision at various stages. 

First in Section \ref{sect:consistency}, we show a consistency check by comparing images taken of the calibration station over the full span of station production. This serves as a check that the stand itself is not deforming over time and is a check of the measurement repeatability. 

In Section \ref{sec:mechresults}, we discuss a few validations of the mechanical measurements. The lengths between the fiducials measured by the laser tracker serves as a check of the lengths measured by the mechanical measurements. In addition, we discuss the repeatability of these mechanical measurements over changes in station orientation. 

In Section \ref{sect:XRot}, we validate the camera-to-camera calibration in the camera array by rotating the calibration station by $\sim 11 \unit{\milli \meter}$
 about the X-axis. We verify
 we observe the expected rigid body rotation in all 36 of the panel fiducials. Here, we compare the images taken before and after the rotation. 

In Section \ref{sect:dukecomp}, we compare the lengths between the three fiducials in a single panel measured by the camera array and the X-ray technique; this is critical for transforming the X-ray dataset into a station-wide coordinate system. 

Finally, in Section \ref{sect:mechradiuscomp} we compare the 3D positions from the camera analysis when using the mechanical measurements to that using tooling ball radius. This is a cross-check of the implementation of the mechanical measurements into the image analysis and a cross check of the radius measurement. 

\subsection{Calibration Station Stability Check}
 \label{sect:consistency}

As a verification of the stability the metrology stand and the calibration station, we periodically imaged the calibration station.  In particular, it is critical to check that the honeycomb itself has not deformed. Any deformation of the honeycomb would result in systematic errors in the fiducial position measurements. A honeycomb deformation could be identified in this study as a non-rigid body transformation between two sets of images of the calibration station, but this also could be the consequence of other issues (e.g. deformation of the calibration station). 

\begin{figure}[htb]
{\centering
\includegraphics[width=0.5\textwidth]{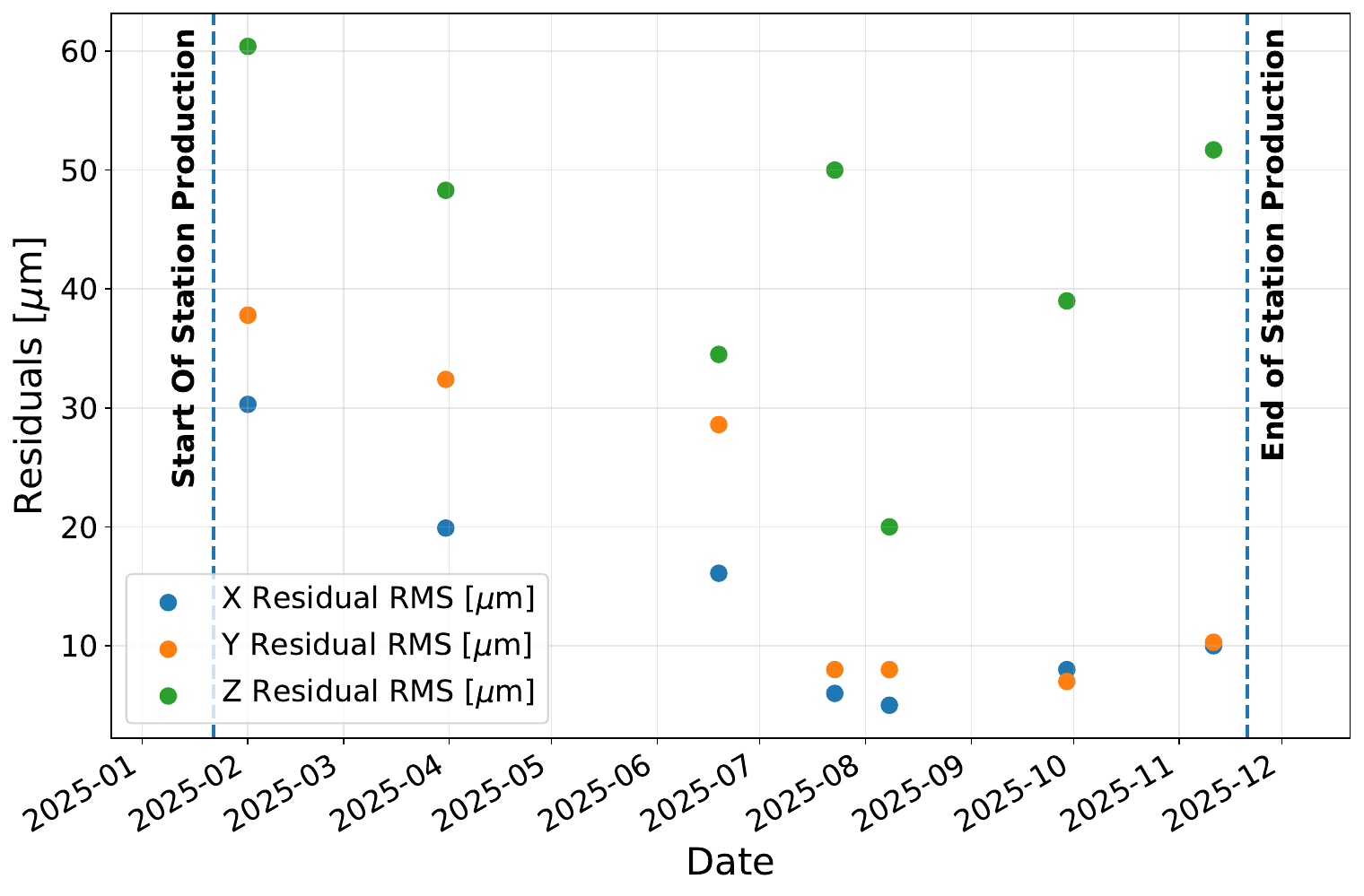}
\caption{The scatter plot shows the XYZ RMS residuals from the fitted rigid body transformation of the calibration station to itself between two sets of images in the metrology stand. The residuals are plotted as a function of date.
\label{CalibStationCompare}
}}\end{figure}

The calibration station was imaged several times over the course of the several months of station production. For each set of images we transform the data into the laser tracker frame and compare this to the reference set of images (taken at the time of the laser tracker survey). The XYZ RMS residuals of the rigid body transformation from the two sets of calibration station image data is shown in Figure \ref{CalibStationCompare}. We observe RMS residuals in XY <  $40 \unit{\micro\meter}$ and $\sim 50 \unit{\micro\meter}$ in Z. This checks that the station itself was not deforming significantly, the camera-camera alignment did not change significantly, and the honeycomb did not deform significantly. 

However, the rigid body fit above did result in translations and rotations. This motion is dominated by X translations from one image set to another at the scale of $<100 \unit{\micro \meter}$. This is expected as the stand is not infinitely rigid. The stand legs move inward (X) due to the weight of the stand, thus moving the station with respect to the honeycomb and the cameras.  

To correct for these stand distortions, we apply the following procedure. For each production station we rely on the single set of laser tracker fiducial positions ($X_{C, LT}$). We take the corrected position of the production station fiducial ($X_{P', LT}$) as $X_{C, LT}$ minus the difference between the production station fiducial position, $X_{P, LT}$, and the calibration station fiducial position in the most recent image set, $X_{C', LT}$, that is: 

$$X_{P', LT} = (X_{P, LT} - X_{C', LT}) + X_{C, LT}$$

As an example, if the stand distorts and all fiducials move in $X_{LT}$, this should be seen as a shift in both $X_{P, LT}$ and $X_{C', LT}$, and the stand distortion cancels out. Note, if the stand deformed in between calibration station image sets, this would result in a misalignment in the station-to-station alignment, but not a panel-panel alignment within a station.

\subsection{Mechanical Measurement Validations}
\label{sec:mechresults}
Here, we discuss two validations of the mechanical measurements. First, we use the laser tracker of the calibration station to make a comparison between the lengths between the fiducials in a triplet measured by the laser tracker to the three mechanical measurements per triplet. This is equivalently the residuals from the fitted transformation from the camera frame to the laser tracker frame for each camera. 

In Table \ref{tab:rms_vs_angle} we tabulate the RMS of the fit for the transformation from the camera frame to the laser tracker frame for each camera including the mechanical measurements. This is effectively the errors in the 3 lengths measured by the laser tracker and measured by the mechanical measurements. The 2nd column is the fitted angle about the Z axis for each camera with respect to the station-wide frame normalized by the first camera. As expected each camera is rotated by $\sim 30$ degrees with respect to the previous. Most of the residuals are quite good with a single outlier with an RMS of $151 \unit{\micro\meter}$.

 In the final column, instead of using the mechanical measurements, we use the tooling ball radius and the 2D position on the camera CCD to yield the lengths between the fiducials. As expected, we achieve a degraded RMS due to the degraded Z resolution from the radius measurement. The fitted angle about the Z axis was the same as that with the mechanical measurements to a precision below 0.1 degrees. 

\begin{table}[htbp]
\centering
\begin{tabular}{
>{\centering\arraybackslash}p{1.7cm}
>{\centering\arraybackslash}p{1.7cm}
>{\centering\arraybackslash}p{1.7cm}
>{\centering\arraybackslash}p{1.4cm}
}
\hline
Camera ID &
Angle about Z axis [deg] &
RMS Mechanical [$\unit{\micro\meter}$] &
RMS Radius [$\unit{\micro\meter}$] \\
\hline
0  &   0.0   & 17.1  & 192.4 \\
1  & -30.0   & 5.0   & 217.7 \\
2  & -60.2   & 9.4   & 152.9 \\
3  & -89.9   & 18.9  & 242.7 \\
4  & -120.1  & 151.2 & 679.8 \\
5  & -150.4  & 22.7  & 90.6  \\
6  & -180.1  & 24.8  & 266.0 \\
7  & -210.3  & 13.5  & 327.4 \\
8  & -239.4  & 5.0   & 266.7 \\
9  & -270.2  & 21.5  & 157.8 \\
10 & 60.1    & 66.7  & 193.1 \\
11 & 29.8    & 9.5   & 585.6 \\
\hline
\end{tabular}

\caption{RMS residuals from the 3 DOF rigid body fit for the transformation from the camera coordinate system to the laser tracker coordinate system using the mechanical measurements or the tooling ball radius.}
\label{tab:rms_vs_angle}
\end{table}

In addition, to avoid any mechanical measurement error due to the dial failing to make good contact with the fiducial or the dial being mis-measured, we made the measurements with the station horizontal on a granite table after assembly and then repeated the same measurements once the detector was vertical (180 degree rotation about the X axis). The average RMS of the difference in the longest mechanical measurement made in the two orientations over all stations is 0.0005" ($0.0127 \unit{\milli \meter}$), and the RMS of the difference in the other two lengths is 0.002" ($0.0508 \unit{\milli \meter}$). 

The longest length measures the length between two fiducials on different panels in the same plane whereas the other two lengths measure the length between two fiducials on the two planes on a station. The panel-to-panel alignment is more rigid than the plane-plane alignment in a station, thus we expect that the longest mechanical measurement should have the best agreement over orientations. To set the scale, an error of 0.0005" ($0.0127 \unit{\milli \meter}$) in the longest length would create a $81 \unit{\micro\meter}$ $Z_{LT}$ error for the triplet of fiducials at that location.  

Comparing the lengths measured in the two orientations indicates that there is not any significant 'local' deformation of the station even as we change orientations (e.g. panel-panel shifts, plane-plane dX,dY,d$\phi$), and shows the repeatability of these measurements.

\subsection{X-Axis Rotation Check}
\label{sect:XRot}
To check the inter-camera calibration, we adjust the angle of the station about the X-axis and checked the observed fiducial motion matches the expected rigid body motion. We create the rotation by shifting the lower kinematic mount Z position. We took images with the camera array before and after. We fit for the rigid body motion between the two sets of images across all cameras. The slot is located at $Y\sim -840 \unit{\milli\meter}$ from the rotation axis and was shifted by $11.074 \unit{\milli \meter}$, therefore creating a rotation of $13.3 \unit{\milli \radian}$.

Using the standard procedure described above, the image data and mechanical measurements are transformed into a station-wide coordinate system before and after the rotation. We fit for the rigid body transformation between the datasets. The fit results in a 13.9 mrad rotation about the X-axis and RMS residuals of 5.6, 8.1, 32.5 $\unit{\micro\meter}$ in X,Y,Z respectively. Note, this uses the same mechanical measurements so this is effectively only error in the 2D position of the tooling balls measured by the camera. This exercise is meant to show that the position resolution from the cameras far exceeds our requirements and the series of transformations moving from the mechanical measurements + 2D tooling ball positions on the camera CCD are working as expected. In addition, the combination of this check and the mechanical measurement validations in the previous section suggests we should be achieving a resolution of $\sim 100 \unit{\micro \meter}$ along the camera axis when using the mechanical measurements.

If instead we use the 2D position of the tooling balls and the measured radius on the CCD, the fit results in a rotation of 14.3 mrad about the X-axis and RMS residuals of 21.5, 37.2, 327.2 $\unit{\micro\meter}$ in X,Y,Z respectively. We observed a Z resolution of $\sim 300 \unit{\micro\meter}$, thus holding a fractional precision of $\sim 0.3/500=0.0006$ or a radius measurement precision of $0.0006*6.35=4 \unit{\micro\meter}$ over the $11 \unit{\milli \meter}$ shift with a 0.5" tooling ball ($6.35 \unit{\milli\meter}$ radius).

\subsection{Transformations From Panel to Station Coordinate System}
\label{sect:dukecomp}
\begin{figure*}[htb]
{\centering
\includegraphics[width=\textwidth]{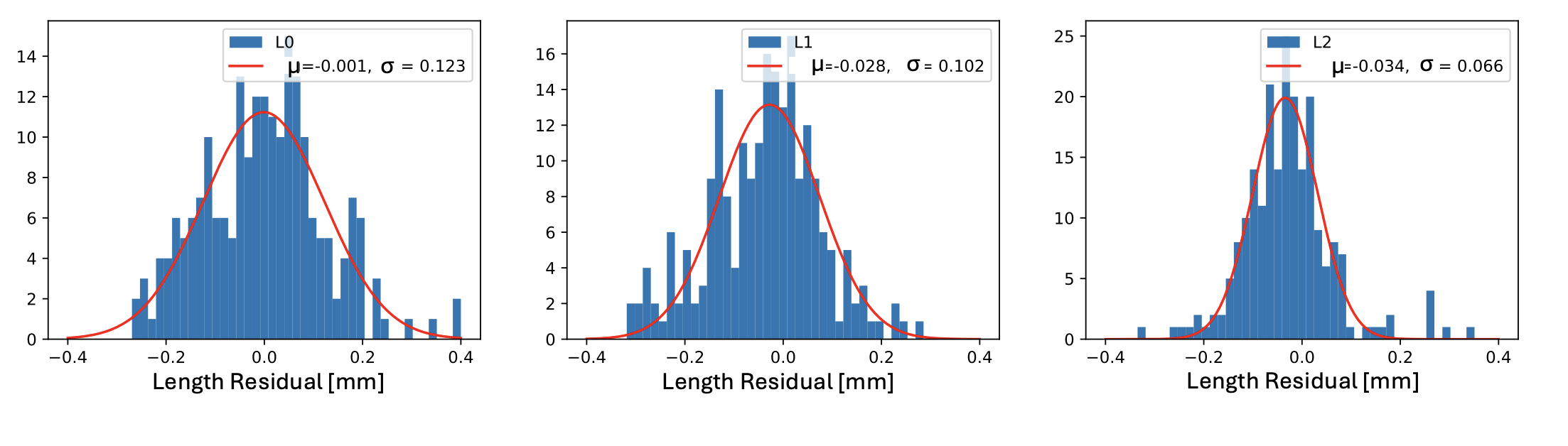}
\caption{Histograms of the lengths between fiducials in a panel measured by the camera array compared to that in the X-ray dataset [$\unit{\milli \meter}$]. The first histogram contains the longest length. 
\label{LengthComp}
}}\end{figure*}
 For each panel, we fit for the rigid body transformation of three fiducials measured in the X-ray survey and those measured in the camera stand in the station-wide coordinate system. 

Here, we have 3x3D positions in the two reference frames. We fit for the 6 parameter transformation from the X-ray coordinate system to the camera data in the station-wide laser tracker frame. The remaining DOF are the lengths between the three fiducials. Note, in the X-ray survey the three fiducials define an XY plane, then in the laser tracker frame they are nearly in the XY plane. This transformation will always return effectively zero residuals in Z (1 or 2 $\unit{\micro \meter}$); a small rotation can always recover any difference. The length errors will be nearly exclusively in the XY plane. For this reason, the comparison between the two datasets does not yield any check of the Z measurements in either dataset. 

In addition, the longest of the three lengths between the fiducials is effectively parallel to the wire axis; therefore it is not relevant (translating the wire along the wire axis). In cases where we visually observed scratching on the spot face or some small amount of epoxy (suggesting maybe more was scratched off), we apply a radially inward correction of $35 \unit{\micro\meter}$. The offset value is based on the differences in the average lengths between fiducials with and without observed epoxy. We show the length differences with this correction applied in Figure \ref{LengthComp}.

\subsection{Comparison of Measurements from Radius and Mechanical Measurements}
\label{sect:mechradiuscomp}

\begin{figure*}[htb]
{\centering
\includegraphics[width=0.9\textwidth]{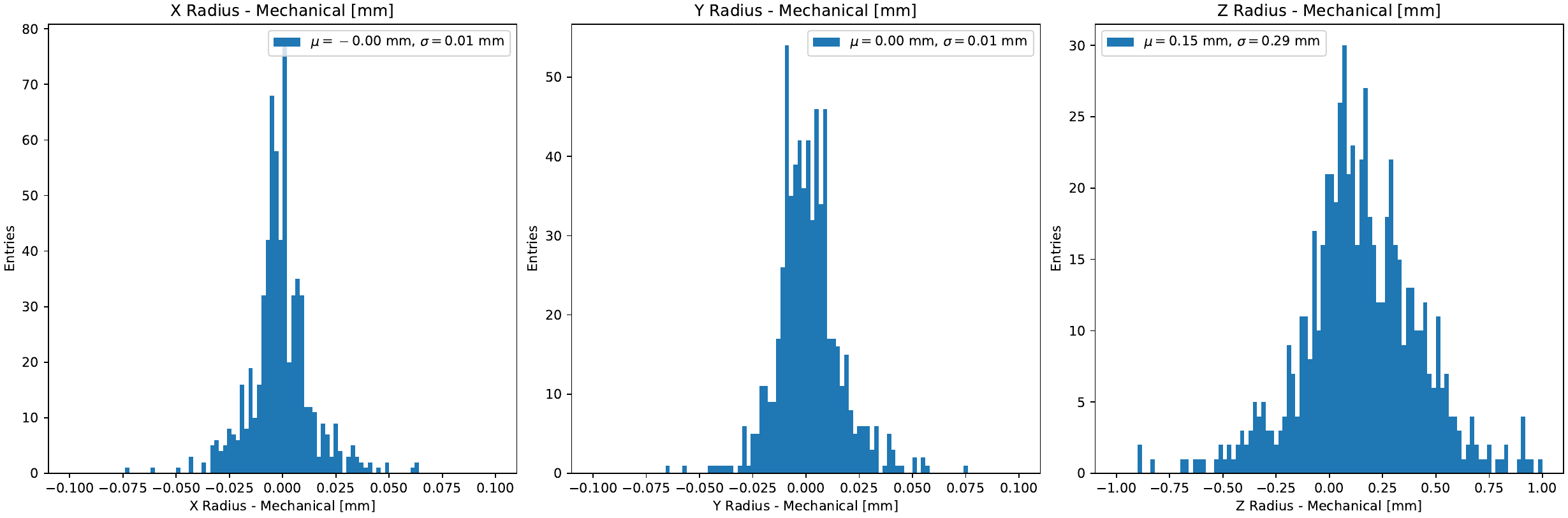}
\caption{Comparison of the $X_{LT}, Y_{LT}, Z_{LT}$
 coordinates over all 648 measurements determined using the radius and the mechanical measurements. The $\sigma $ is an RMS. \label{Radius_Mechanical}
}}\end{figure*}

In this section we compare the XYZ positions found using the two techniques to measure the position along the camera axis: the tooling ball radius and the mechanical measurements. This comparison is shown in Figure \ref{Radius_Mechanical}. As expected, there is a negligible difference in the observed $X_{LT}, Y_{LT}$ positions. 

In the $Z_{LT}$ comparison, we observe a mean difference of $\sim 150 \unit{\micro\meter}$, which could be due to a difference in the tooling ball radius with respect to the nominal listed from the manufacturer. As noted in Section \ref{sect:fids}, the tooling ball radius tolerance is $\sim 5\unit{\micro\meter}$. A systematic error of $5 \unit{\micro \meter}$ in the tooling ball radius scaled to the Z distance results in a Z error of $393.7\unit{\micro\meter}$. If all tooling balls were made in the same batch, this would systematically shift all fiducial positions in $Z_{LT}$. The $Z_{LT}$ comparison has a RMS $=290 \unit{\micro\meter}$. We expect a vast majority of this is error to be from the radius measurement error and is at the level of expected resolution.

\section{Discussion and Conclusions}

\subsection{Results Summary}
In this paper, we describe a metrology technique to determine the relative alignment of the panels in the Mu2e tracker. All 18 stations were measured using a camera array with 15 cameras. The cameras provide 3D reconstruction of the tooling ball positions. We found that the tooling ball radius alone does not yield acceptable position resolution along the camera axis, but this may be acceptable in similar applications. This resolution was improved by supplementing the camera analysis with 'local' mechanical measurements made for each triplet of tooling balls observed by a camera. 

In Section \ref{sect:Results}, we showed we are able to measure a stable position of the tooling balls in the calibration station measured over $\sim6$ months. We then verified the consistency of the local mechanical measurements over different station orientations. 

We presented the results of the transformation from each local camera coordinate system to a station-wide coordinate system; this was checked using both the mechanical measurements and the measured radius of the tooling balls. 

We then used these transformations to measure a large ($>1 $cm) motion via a rotation about the X-axis. We observed changes in the position of the tooling balls consistent with rigid body motion at the level of $<10 \unit{\micro\meter}$ in $X_{Mu2e}$ and $Y_{Mu2e}$ and an RMS of $32 \unit{\micro\meter}$ in $Z_{Mu2e}$. 

We then used the optically-measured position of the panels in the production stations and the position of the panels measured by the X-ray scan to fit for the transformation from the local panel coordinate system to a station-wide coordinate system. The verification was the consistency of the lengths between the three fiducials in a panel measured in the X-ray scan and the camera techniques. Unfortunately, this was clouded by the spot faces on the panels having varying levels of epoxy and scratches. For some panels, epoxy was scratched off between the X-ray measurement and the camera measurement. Nonetheless, we observe differences in the lengths between the fiducials of the order $100 \unit{\micro\meter}$ with one of the three lengths having a fitted Gaussian width of $\sigma=66\unit{\micro\meter}$; it seems this pair of fiducials may be less affected by the epoxy/scratching. Here, the lengths are of the order $\sim 1000 \unit{\milli \meter}$ (longest of $1379 \unit{\milli \meter}$ and the other two of $784 \unit{\milli \meter}$) so our fractional precision is a part in $\sim10,000$. 

This precision is achieved by measuring the position of the calibration station, then only relying on the cameras to measure small changes in the positions. A key to reaching this precision is the stability of the camera-camera alignment, in part due to the rigidity of the aluminum honeycomb and the honeycomb's kinematic mount. The mechanical measurements were added to achieve a significantly more precise measurement along the camera axis. 

We detailed the procedure to transform the data from the metrology stand into a global tracker frame coordinate system and how to deal with potential frame distortions from e.g. transporting the tracker.

\subsection{Usage in a Different Experiment}
The cameras allowed for rapid precision measurements for the 18 stations, with the metrology taking O(1 hour) per station. The procedure yielded high precision measurements at an affordable cost. 

This technique could be applied in similar instances with a series of planar detectors where the relative alignment of the sub-modules in a plane are measured optically, then placed in the final detector using kinematic mounts. 

\begin{figure}[htb]
{\centering
\includegraphics[width=0.49\textwidth]{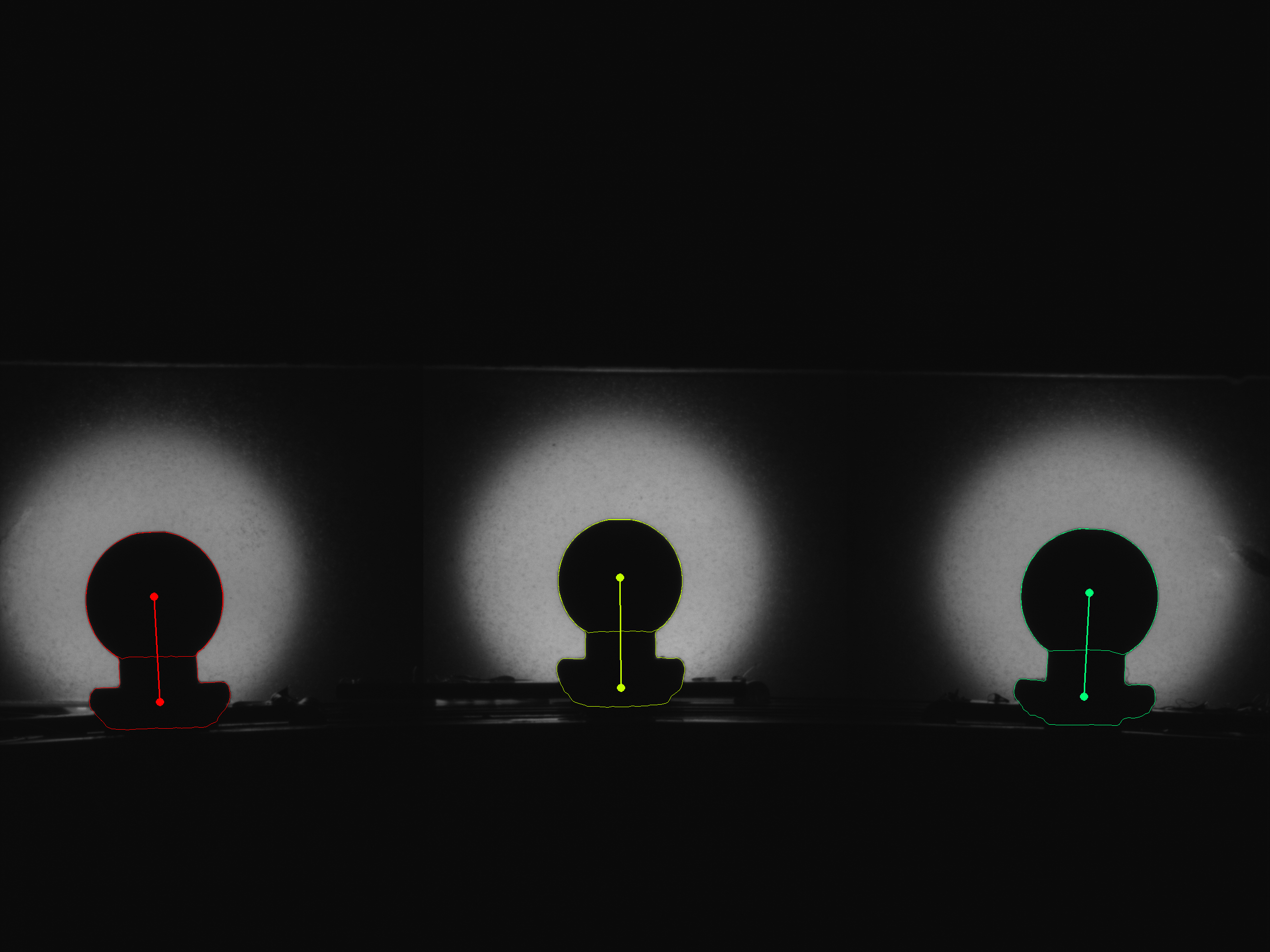}
\caption{Image of a triplet of tooling ball with Segment Anything\cite{kirillov2023segment} found masks plus the tooling ball directional vector from the masks.   
\label{SegmentAnything}
}}\end{figure}

In addition, instead of only measuring tooling balls, one could envision a situation where there was no X-ray scan (or analogous scan) of the wires/straws in a panel. Instead, the straws or other detector components, e.g. SiPMs could be measured optically by the camera array. 
With the improvements in AI imaging algorithms, more complex detector component contours could be more easily optically measured. To see what current AI algorithms could detect, we implemented Meta's Segment Anything algorithm\cite{kirillov2023segment} on the images. We tested if 'out-of-the-box' AI could find the tooling balls and whether any additional features (e.g. the tooling ball stem) could be found. The algorithm is pre-trained on a large set of images to find relevant masks. This out-of-the-box AI with mask selection based on area (avoiding the LEDs and small masks found), resulted in masks for the sphere on each tooling ball and the full tooling ball (ball+stem+base). Using the two masks per tooling ball one can determine, for example, the direction vector of the tooling ball. This direction vector and the contours are shown in Figure \ref{SegmentAnything}. While this is not unique to AI and could be determined using analysis like OpenCV, it is interesting to see that these features could be found with no training and minimal selection.

\subsection{Conclusions} 

The paper discusses a photogrammetry approach to measure the panel-panel alignment in the Mu2e tracker using an array of 15 cameras mounted on a single aluminum honeycomb.  Combining these measurements with a previous X-ray survey of the straw tube positions yields the position of the wires and straws in a station-wide coordinate system. 

The technique provides rapid position measurements over a $\sim1.6$m span with a precision of $\sim 30 \unit{\micro\meter}$ in the camera transverse directions. Along the camera axis, the resolution is $< 100 \unit{\micro\meter}$ when using the local mechanical measurements and $\sim 300 \unit{\micro\meter}$ using the imaged radius. 

In addition, the use of kinematic mounts and precision fiducials on the tracker frame allow for these single-station measurements to be transformed into a global tracker frame coordinate system.  

\section{Acknowledgments}

Thank you to the full tracker team for their support in this project. Particular thanks to Kourosh Taheri, Phillip Cowan, Otto Alvarez for the required machining, and Adrian Marquez for the help with tabulating panel IDs, spot face checks, and mechanical measurements. In addition, we would like to thank Seog Oh and Chiho Wang for their help with comparing the camera measurements to the X-ray dataset.

We are grateful for the vital contributions of the Fermilab staff and the technical staff of the participating institutions. This work was supported by the US Department of Energy; the Istituto Nazionale di Fisica Nucleare, Italy; the Science and Technology Facilities Council, UK; the Royal Society, UK; the Leverhulme Trust, UK; the Ministry of Education and Science, Russian Federation; the National Science Foundation, USA; the National Science Foundation, China; the Helmholtz Association, Germany; the Julian Schwinger Foundation, USA; the EU Horizon 2020 Research and Innovation Program under the Marie Sk\l{}odowska-Curie Grant Agreement Nos.\ 858199, 101003460, 101006726 and 101230211; and the Horizon Europe Research and Innovation Program under the Marie Sk\l{}odowska-Curie Grant Agreement No.\ 101234557. This document was prepared by members of the Mu2e Collaboration using the resources of the Fermi National Accelerator Laboratory (Fermilab), a U.S.\ Department of Energy, Office of Science, Office of High Energy and Nuclear Physics HEP User Facility. Fermilab is managed by FermiForward Discovery Group, LLC, acting under Contract No.\ 89243024CSC000002.

\bibliographystyle{unsrt}   % or plain, alpha, ieeetr, apsrev4-2, etc.
\bibliography{references}

\end{document}